\documentclass[preprint]{aastex}
\usepackage{graphics}
\usepackage{ifthen}

\shorttitle{K Band Spectroscopy of ULIRGs}
\shortauthors{Murphy et al.}
\slugcomment{To appear in the \emph{Astronomical Journal}}

\newcommand{\mycaption}[2]{ %
  \vspace{0pt}\sbox{\tempbox}{\small #1: #2}%
  \ifthenelse{\lengthtest{\wd\tempbox > \linewidth}} %
  {\small #1: #2\par}
  {\begin{center}\small #1: #2\end{center}}%
}
\newsavebox{\tempbox}

\begin{document}

\title{\( K \) Band Spectroscopy of Ultraluminous Infrared Galaxies: The 2~Jy Sample}

\author{T. W. Murphy, Jr., B. T. Soifer\altaffilmark{1}, K. Matthews}

\affil{Palomar Observatory, California Institute of Technology, 320--47, Pasadena,
CA 91125}

\email{tmurphy@mop.caltech.edu, bts@mop.caltech.edu, kym@caltech.edu}

\author{L. Armus }

\affil{SIRTF Science Center, California Institute of Technology, 314--6, Pasadena,
CA 91125}

\and{}

\author{J. R. Kiger}

\affil{Center for Space Research, Massachusetts Institute of Technology, 70 Vassar
Street, Building 37, Cambridge, MA 02139}

\altaffiltext{1}{Also at SIRTF Science Center, California Institute of Technology, 314--6, Pasadena, CA 91125}

\begin{abstract}
We present near-infrared spectroscopy for a complete sample of 33 ultraluminous
infrared galaxies at a resolution of \( R\approx 1000 \). Most of the wavelength
range from 1.80--2.20 \( \mu  \)m in the rest frame is covered, including the
Pa\( \alpha  \) and Br\( \gamma  \) hydrogen recombination lines, and the
molecular hydrogen vibration-rotation 1--0 S(1) and S(3) lines. Other species,
such as He~I, {[}Fe~II{]}, and {[}Si~VI{]} appear in the spectra as well,
in addition to a number of weaker molecular hydrogen lines. Nuclear extractions
for each of the individual galaxies are presented here, along with spectra of
secondary nuclei, where available. The Pa\( \alpha  \) emission is seen to
be highly concentrated on the nuclei, typically with very little emision extending
beyond a radius of 1~kpc.

This survey was carried out primarily to search for signatures of active nuclei
via velocity-broadened hydrogen recombination or the presence of the {[}Si~VI{]}
coronal line. These signatures are rare in the present sample, occurring in
only two of the 33 galaxies. The extinction to the hydrogen recombination lines
is investigated via the Pa\( \alpha / \)Br\( \gamma  \) line ratio. It is
found that visual extinctions to the nuclei in excess of 10 magnitudes are relatively
common among ULIRGs, and that visual extinctions greater than 25 mag are necessary
to conceal a QSO emitting half the total bolometric luminosity. The ionized
hydrogen regions in many ULIRGs are sufficiently obscured that dust-enshrouded
active galactic nuclei would remain hidden at 2 \( \mu  \)m at the current
level of sensitivity. The vibration-rotation lines of molecular hydrogen appear
to be predominantly thermal in origin, with effective temperatures generally
around 2200 K. The relative nuclear velocities between double nucleus ULIRGs
are investigated, through which it is inferred that the maximum deprojected
velocity difference is \( \sim 200 \) km~s\( ^{-1} \). This figure is lower
than the velocities predicted by physical models of strong interactions/mergers
of large, gas-rich galaxies.
\end{abstract}


\section{Introduction}

Understanding the nature of the ultimate power source in ultraluminous infrared
galaxies (ULIRGs) has been a primary driver for numerous scientific investigations
ever since the discovery in the 1980's that ULIRGs represent the most luminous
class of galaxies in the local universe. With total power outputs comparable
to those of quasars, characterizing the source of power is of fundamental importance
in understanding the composition and evolution of normal galaxies. ULIRGs are
found to share the common trait that virtually all have recently or are currently
experiencing major encounters with other galaxies \citep{twm96,clements,sanders,carico}.
Encounters like these may play a significant role in the process of building
normal galaxies. With the discovery of faint sources at sub-millimeter wavelengths
having characteristic spectral energy distributions like that typically found
in ULIRGs (e.g., Arp~220) and redshifts \( z\geq 2 \), the importance of understanding
these systems and their role in the evolution of normal galaxies has been heightened.

The present survey was designed to look for near-infrared spectral signatures
of obscured active galactic nuclei (AGN) in ULIRGs, either by the presence of
velocity-broadened Pa\( \alpha  \) emission or by the appearance of the high
excitation {[}\ion{Si}{6}{]} coronal line. A summary of the conclusions of this
survey, along with a median ULIRG near-infrared spectrum, are presented in \citet{twm99}.
Only two of the 33 sample galaxies show clear indications for AGN activity,
with the rest appearing to be dominated by star formation processes. Either
the AGN phenomenon is relatively rare among ULIRGs, or the extinction at 2 \( \mu  \)m
is still too high to permit a proper assessment. Mid infrared studies \citep[e.g.,][]{genzel,rig}
favor the former conclusion, though it is found in this study that the extinction
at 2 \( \mu  \)m may often be greater than four magnitudes, resulting in over
one magnitude of extinction at mid-infrared wavelengths.

This paper presents the individual nuclear spectra for the sample galaxies,
for both the primary and secondary nuclei where possible. Two-dimensional spectra
are also presented to give a qualitative view of the line emission distributions
typically found in ULIRGs. Some discussion of extinction measures, molecular
hydrogen excitation mechanisms, and a summary of typical line ratios found in
ULIRGs is included, as is a discussion of the velocity differences found among
double nucleus systems.

\section{The Sample}

The sample of 33 ULIRGs in this survey originates from the 2Jy sample of \emph{IRAS}
(Infrared Astronomical Satellite) galaxies defined by \citet{str90,str92}. The
galaxies in the present sample are chosen to meet the following criteria:

\begin{enumerate}
\item 60 \( \mu  \)m flux density, \( F_{\nu } \)(60 \( \mu  \)m) \( > \) 1.94
Jy
\item Moderate or high quality IRAS flux measurements at 60 and 100 \( \mu  \)m \citep[as defined in the][]{iraspsc}
\item \( F_{\nu }^{2} \)(60 \( \mu  \)m)\( >F_{\nu } \)(12 \( \mu  \)m)\( \times F_{\nu } \)(25
\( \mu  \)m), where \( F_{\nu } \)(\( \lambda  \)) is the \emph{IRAS} flux
density
\item Far-infrared luminosity \( L_{FIR}>5\times 10^{11}L_{\odot } \)
\item Absolute Galactic latitude, \( \left| b\right| >5^{\circ } \)
\item Declination, \( \delta >-35^{\circ } \)
\item Redshift, \( 0.055<z<0.108 \)
\end{enumerate}
The first six conditions are those used to define the northern 2 Jy ULIRG sample
in \citet{twm96}, such that the current sample is a subset of this work. The
seventh condition places both the Pa\( \alpha  \) and Br\( \gamma  \) lines
in the \( K \) band atmospheric window. The far-infrared luminosity is computed
from the IRAS flux densities by the following conversion:\[
L_{FIR}\equiv 3.86\times 10^{5}D^{2}\left[ 2.58F_{\nu }(60\, \mu \rm m)+F_{\nu }(100\, \mu \rm m)\right] L_{\odot },\]
\citep[][p. X-13]{iraspsc} where \( D \) represents the luminosity distance
to the source in Mpc. 

Due to the redshift constraint, the current sample approximately represents
a complete volume-limited sample, because even the galaxies with the minimum
60 \( \mu  \)m flux density are luminous enough to meet the sample criteria
at all allowed redshifts. At the outer redshift limit of the survey, corresponding
to a luminosity distance of 455 Mpc using \( H_{0}=75 \) km~s\( ^{-1} \)~Mpc\( ^{-1} \)
and \( q_{0}=0 \) (as is assumed throughout), the minimum 60 \( \mu  \)m flux
density of 1.94 Jy by itself accounts for \( 4\times 10^{11}L_{\odot } \) of
the total far-infrared luminosity. To make up the remaining \( 10^{11}L_{\odot } \)
necessary for sample inclusion, the 100 \( \mu  \)m flux density needs only
be greater than 1.25 Jy, implying a maximum \( F_{\nu } \)(60 \( \mu  \)m)\( /F_{\nu } \)(100
\( \mu  \)m) ratio of 1.55. ULIRGs very rarely have 60 \( \mu  \)m to 100
\( \mu  \)m flux density ratios this high, such that the volume limit imposed
by the upper redshift cutoff is more stringent than the flux density cutoff,
resulting in a sample that is mostly volume-limited rather than flux-limited.
Taking the 64 ULIRGs defined in the northern 2 Jy sample of \citet{twm96}, one
finds an average \( F_{\nu } \)(60 \( \mu  \)m)\( /F_{\nu } \)(100 \( \mu  \)m)
ratio of \( 0.90\pm 0.22 \), with a median value of 0.86 and a maximum of 1.64
(in IRAS~08572\( + \)3915), which is the only ratio exceeding the maximum
of 1.55 compatible with the volume-limited case. The ULIRG from the northern
2 Jy sample with the second highest ratio is IRAS~05246\( + \)0103, with \( F_{\nu } \)(60
\( \mu  \)m)\( /F_{\nu } \)(100 \( \mu  \)m)\( =1.31 \). Both of these galaxies
are contained in the present sample. The redshift limit excludes many of the
highest luminosity ULIRGs, as these are most likely found at larger distances.
For instance, Mrk~1014---a quasar and a ULIRG---would be a member of the sample
if not for the upper redshift limit.

Of the 35 ULIRGs in the redshift survey of \citet{str92} satisfying the above
conditions, two are not included here for the following reasons. The spectra
of IRAS~21396\( + \)3623 show this galaxy to be at a redshift of \( z=0.1493 \),
rather than the previously reported \( z=0.097 \). IRAS~19297\( - \)0406
was free of any discernible lines, which may be the result of improper pointing,
producing the spectrum of a nearby star in the very crowded field. Therefore
the entire sample contains 33 galaxies, and is very nearly complete in the volume-limited
sense. A list of the sample galaxies appears in Table~\ref{tab:sample}.

\begin{deluxetable}{lccc} 
\tabletypesize{\footnotesize}
\tablewidth{0pt}
\tablecaption{Spectroscopic Survey Sample\label{tab:sample}}
\tablehead{\colhead{Galaxy Name} & \colhead{$cz$} & \colhead{$L_{IR}$} & \colhead{Separation\tablenotemark{a}} \\
& \colhead{(km s$^{-1}$)} & \colhead{($10^xL_\odot$)} & \colhead{(arcsec)} \\ }
\startdata 

IRAS 00262$+$4251 & 29205 & 12.08 & S \\

IRAS 01521$+$5224 & 23959 & 11.95 & 5.5 \\

IRAS 04232$+$1436 & 23640 & 11.99 & S \\

IRAS 05246$+$0103\tablenotemark{b} & 29105 & 12.05 & 6.0 \\

IRAS 08030$+$5243 & 24946 & 11.97 & S \\

IRAS 08311$-$2459\tablenotemark{b} & 30167 & 12.40 & S \\

IRAS 08344$+$5105 & 28999 & 11.94 & S \\

IRAS 08572$+$3915\tablenotemark{b,c} & 17491 & 12.08 & 5.6 \\

IRAS 09061$-$1248 & 22014 & 11.97 & 5.6 \\

IRAS 09111$-$1007 & 16247 & 11.98 & (37) \\

IRAS 09583$+$4714 & 25748 & 11.98 & 25.5 \\

IRAS 10035$+$4852 & 19449 & 11.93 & 9.8 \\

IRAS 10190$+$1322 & 22898 & 11.98 & 4.0 \\

IRAS 10494$+$4424 & 27671 & 12.15 & S \\

IRAS 11095$-$0238 & 31936 & 12.20 & S \\

IRAS 12112$+$0305\tablenotemark{c} & 22009 & 12.27 & 2.9 \\

IRAS 14348$-$1447\tablenotemark{c} & 24798 & 12.28 & 3.3 \\

IRAS 14352$-$1954 & 26938 & 11.95 & S \\

IRAS 14394$+$5332 & 31401 & 12.04 & (28) \\

IRAS 15245$+$1019 & 22629 & 11.96 & 2.6 \\

IRAS 15250$+$3609\tablenotemark{c} & 16515 & 11.99 & S \\

IRAS 15462$-$0450 & 30060 & 12.16 & S \\

IRAS 16487$+$5447 & 31110 & 12.12 & 3.1 \\

IRAS 17028$+$5817 & 31805 & 12.11 & 13.0 \\

IRAS 18470$+$3233 & 23520 & 12.02 &(7.0) \\

IRAS 19458$+$0944 & 29983 & 12.36 & S \\

IRAS 20046$-$0623 & 25286 & 12.02 & S \\

IRAS 20087$-$0308 & 31613 & 12.39 & S \\

IRAS 20414$-$1651 & 26061 & 12.19 & S \\

IRAS 21504$-$0628 & 23343 & 11.94 & S \\

IRAS 22491$-$1808\tablenotemark{c} & 23264 & 12.12 & 1.6 \\

IRAS 23327$+$2913 & 32225 & 12.03 & 12.5 \\

IRAS 23365$+$3604 & 19305 & 12.13 & S \\
 \enddata
\tablenotetext{a}{Nuclear separation is given where applicable. S denotes a single nucleus. Numbers in parentheses denote a double nucleus system for which only the primary nucleus spectrum is presented, generally because the separation exceeds the slit length.}
\tablenotetext{b}{Galaxies with ``warm'' infrared colors: $F_\nu$(25 $\mu$m)$/F_\nu$(60 $\mu$m)$> 0.2$.} 
\tablenotetext{c}{Contained in the Bright Galaxy Sample (BGS) of \citet{btsbgs}.}
\end{deluxetable}

\section{Observations and Data Reduction}

Observations were conducted using the Palomar longslit infrared spectrograph
\citep{larkin}, operating on the 200-inch Hale Telescope and employing a 256\( \times  \)256
HgCdTe (NICMOS~3) array. Observations were made over a time period beginning
in 1995 July, and ending in 1997 December. Each object was observed in three
grating settings, with each spanning \( \sim 0.12 \) \( \mu  \)m at a scale
of \( \sim 0.0006 \) \( \mu  \)m~pixel\( ^{-1} \). Typical slit widths of
\( \sim  \)3.75 pixels, or 0\farcs 625, resulted in spectra with resolutions
of \( R\equiv \lambda /\Delta \lambda \approx 1000 \), corresponding to \( \sim  \)300
km~s\( ^{-1} \). The grating settings were chosen to cover the Pa\( \alpha  \)
line at 1.8751 \( \mu  \)m, the suite of Br\( \delta  \), H\( _{2} \)~1--0
S(3), and {[}\ion{Si}{6}{]} lines centered at 1.954 \( \mu  \)m, and the H\( _{2} \)~1--0
S(1) and Br\( \gamma  \) lines centered at 2.14 \( \mu  \)m. The slit was
generally rotated such that spectra of both primary and secondary nuclei were
obtained simultaneously when possible, or otherwise to coincide with the major
axis of the galaxy, where evident.

Observational details are provided for each source in Table~\ref{tab:obs}.
Individual exposure times were 300 s, with total integration times generally
around 1800 s. In all cases, the galaxy was dithered between two positions on
the slit, generally \( \sim 20'' \) apart, with a smaller scale dither pattern
employed about these points to eliminate the effects of static bad pixels. In
this way, the sky integrations were obtained simultaneously, with the effective
sky position alternating by \( 20'' \) to either side. Wavelength calibration
was provided either by the atmospheric OH airglow spectrum, or by arc lamp spectra
for the H\( _{2} \)~1--0 S(1)\( + \)Br\( \gamma  \) grating settings, where
the OH lines become unavailable. Wavelengths reported here are in air units,
with the OH wavelength data coming from \citet{oo}. Atmospheric transmission
variations as a function of wavelength were compensated via observations of
the nearly featureless spectra of G dwarf stars at similar airmass, and similar
telescope pointing, when possible. The G star observations were performed either
immediately prior to or following each galaxy observation. Stellar types and
airmasses of the calibrators are listed for each object in Table~\ref{tab:obs}.
The light from the calibrator star was made to uniformly fill the slit aperture
by chopping the telescope secondary mirror in a triangle-wave pattern. As such,
the atmospheric calibrators perform the dual function of operating as the flat
field for the spectral observations. Details about the treatment of the spectral
calibration appear below. The calibration star observation for IRAS~12112\( + \)0305
was compromised beyond repair, so the spectra for this galaxy were divided by
an atmospheric template. 

Each spectrum is reduced pair-wise in the following manner. Pairs containing
spectra at the two opposite slit positions are subtracted, with cosmic rays
and static bad pixels interpolated. The data are divided by the G star spectrum
and multiplied by a Planck function, described in more detail below. A wavelength
reference image is produced using either a median combination of the data frames
to produce an OH airglow spectrum, or by using a combination of noble gas arc
lamp spectra. The wavelength calibration image is spatially and spectrally rectified
onto a rectilinear grid using predetermined maps of the distortion in each of
these directions. A cubic function of fixed cubic coefficient and zero quadratic
coefficient (determined empirically to fit the distortion well) is fit to the
wavelength reference, providing a mapping from pixel coordinates to linear wavelength
coordinates. This mapping is combined with the predetermined distortion maps,
and then applied in a single interpolation to the data frames. Once rectified,
the pairs are combined into a single spectrum, using Gaussian centroids for
the spatial registration, or known pixel offsets for galaxies having diffuse
or faint continua. This process removes any residual sky subtraction signal,
as one effectively subtracts the simultaneous sky at the alternate slit position
in the process. The noise at each pixel in the two-dimensional spectrum is computed
along with the spectral data by quadrature combination of read noise and Poisson
shot noise based on signal level for each pixel. Pixel replacement, atmospheric
calibrator division, and interpolation in the data frames are also properly
treated in the noise frames throughout the process. The same extractions applied
to the spectral data are performed on the noise image, producing a one-dimensional
noise array along with the extracted spectrum.

The seeing at the time of observation was assessed by taking short guided exposures---typically
10 s---of nearby field stars through the wide-open slit of the spectrograph,
with the spectrograph operating in imaging mode. In this way, the point spread
function (PSF) of the telescope-plus-atmosphere during the spectral observations
could be estimated. These PSF exposures were typically taken 3--4 times during
an observation, usually at the beginnings and ends of the 30--40 minute intervals
spent at each grating setting. The full-width at half-maximum (FWHM) along the
slit direction was measured for each PSF image, and the nuclear extractions
of the spectra were based on this measurement. In each case, the PSF FWHM was
rounded up to an integral number of pixels, and extractions matching these integer
widths were performed. The centers of these apertures were not constrained,
however, such that fractional pixels were summed in the production of the nuclear
spectra. The widths of the extractions, in arcseconds, are listed in Table~\ref{tab:obs}.

\begin{deluxetable}{lrcccccc}
\tabletypesize{\scriptsize}
\tablewidth{0pt}
\tablecaption{Observational Parameters\label{tab:obs}}
\tablehead{\colhead{Galaxy} & \colhead{Grating} & \colhead{Date} & \colhead{Integration} & \colhead{Slit} & \colhead{Atmos.} & \colhead{Obj./Cal.} & \colhead{Extraction }\\
& \colhead{Setting} & & \colhead{(sec)} & \colhead{P.A.} & \colhead{Calib.} & \colhead{Airmass} & \colhead{($''$)} \\ }
\startdata
IRAS 00262$+$4251 & Pa$\alpha$: & 3 Jan 1996 & 1800 & \phn\phn$0^\circ$ & G1.5V & 1.02/1.05 & 1.5 \\*

~& [Si VI]: & & 1800 & & & 1.11/1.04 & 1.7 \\*

~& H$_2+$Br$\gamma$: & & 1800 & & & 1.05/1.05 & 1.5 \\

IRAS 01521$+$5224 & Pa$\alpha$: & 2 Jan 1996 & 1800 & \phn$26^\circ$ & G5Vb & 1.06/1.09 & 2.0 \\*

~& [Si VI]: & & 1800 & & & 1.06/1.08 & 1.8 \\*

~& H$_2+$Br$\gamma$: & & 1800 & & & 1.07/1.08 & 1.7 \\

IRAS 04232$+$1436 & Pa$\alpha$: & 26 Nov 1996 & 1800 & \phn\phn$5^\circ$ & G0V & 1.07/1.04 & 0.7 \\*

~& [Si VI]: & & 1800 & & & 1.07/1.06 & 0.8 \\*

~& H$_2+$Br$\gamma$: & & 1800 & & & 1.06/1.06 & 0.7 \\

IRAS 05246$+$0103 & Pa$\alpha$: & 5 Dec 1996 & 1200 & $109^\circ$ & G1V & 1.19/1.18 & 0.8 \\*

~& [Si VI]: & & 1800 & & & 1.21/1.21 & 0.8 \\*

~& H$_2+$Br$\gamma$: & & 1800 & & & 1.18/1.17 & 0.7 \\

IRAS 08030$+$5243 & Pa$\alpha$: & 31 Dec 1995 & 1800 & $151^\circ$ & G5V & 1.07/1.07 & 1.5 \\*

~& [Si VI]: & & 1800 & & & 1.07/1.06 & 1.3 \\*

~& H$_2+$Br$\gamma$: & & 1800 & & & 1.06/1.07 & 1.3 \\

IRAS 08311$-$2459 & Pa$\alpha$: & 8 Apr 1996 & 1800 & \phn\phn$0^\circ$ & G3V & 1.91/1.99 & 1.7 \\*

~& [Si VI]: & & 1800 & & & 1.92/1.82 & 1.7 \\*

~& H$_2+$Br$\gamma$: & & 1200 & & & 1.94/1.93 & 1.0 \\

IRAS 08344$+$5105 & Pa$\alpha$: & 3 Jan 1996 & 1800 & \phn\phn$0^\circ$ & G0V & 1.09/1.10 & 1.7 \\*

~& [Si VI]: & & 1800 & & & 1.05/1.09 & 1.5 \\*

~& H$_2+$Br$\gamma$: & & 1800 & & & 1.07/1.09 & 1.5 \\

IRAS 08572$+$3915 & Pa$\alpha$: & 25 Nov 1996 & 2400 & $150^\circ$ & G0V & 1.01/1.01 & 0.8 \\*

~& [Si VI]: & & 1200 & & & 1.03/1.02 & 0.7 \\*

~& H$_2+$Br$\gamma$: & & 1800 & & & 1.01/1.01 & 0.7 \\

IRAS 09061$-$1248 & Pa$\alpha$: & 28 Nov 1996 & 1800 & \phn$21^\circ$ & G0V & 1.49/1.51 & 1.5 \\*

~& [Si VI]: & & 1800 & & & 1.47/1.43 & 2.0 \\*

~& H$_2+$Br$\gamma$: & & 1800 & & & 1.45/1.46 & 1.8 \\

IRAS 09111$-$1007 & Pa$\alpha$: & 11 Feb 1997 & 1800 & \phn$40^\circ$ & G2V & 1.39/1.45 & 1.7 \\*

~& [Si VI]: & & 1200 & & & 1.42/1.39 & 1.5 \\*

~& H$_2+$Br$\gamma$: & & 1800 & & & 1.39/1.41 & 1.7 \\

IRAS 09583$+$4714 & Pa$\alpha$: & 3 Jan 1996 & 1800 & 56\fdg 5 & G0V & 1.03/1.03 & 1.5 \\*

~& [Si VI]: & & 1800 & 56\fdg 5 & & 1.03/1.04 & 1.2 \\*

~& H$_2+$Br$\gamma$: & 31 Dec 1995 & 1800 & $125^\circ$ & & 1.04/1.03 & 0.8 \\

IRAS 10035$+$4852 & Pa$\alpha$: & 22 Feb 1997 & 1200 & \phn$33^\circ$ & G0V & 1.04/1.05 & 1.0 \\*

~& [Si VI]: & 23 May 1997 & 1800 & & G1V & 1.21/1.15 & 0.8 \\*

~& H$_2+$Br$\gamma$: & & 1800 & & F9V & 1.09/1.06 & 0.7 \\

IRAS 10190$+$1322 & Pa$\alpha$: & 22 May 1997 & 1800 & \phn$65^\circ$ & G0V & 1.15/1.16 & 1.0 \\*

~& [Si VI]: & & 1800 & & G1V & 1.26/1.23 & 1.2 \\*

~& H$_2+$Br$\gamma$: & & 1800 & & G0V & 1.44/1.42 & 1.0 \\

IRAS 10494$+$4424 & Pa$\alpha$: & 22 Feb 1997 & 1800 & $148^\circ$ & G2V & 1.03/1.02 & 1.0 \\*

~& [Si VI]: & & 1200 & & G0Vs & 1.15/1.10 & 0.8 \\*

~& H$_2+$Br$\gamma$: & & 1800 & & & 1.09/1.10 & 0.7 \\

IRAS 11095$-$0238 & Pa$\alpha$: & 6 May 1996 & 1800 & \phn$38^\circ$ & G1V & 1.24/1.26 & 1.0 \\*

~& [Si VI]: & & 1800 & & & 1.29/1.31 & 1.0 \\*

~& H$_2+$Br$\gamma$: & & 1800 & & & 1.25/1.29 & 0.8 \\

IRAS 12112$+$0305 & Pa$\alpha$: & 5 May 1996 & 1800 & \phn$34^\circ$ & \nodata & 1.09/\nodata & 1.0 \\*

~& [Si VI]: & & 1800 & & & 1.05/\nodata & 0.8 \\*

~& H$_2+$Br$\gamma$: & & 1800 & & & 1.07/\nodata & 1.0 \\

IRAS 14348$-$1447 & Pa$\alpha$: & 5 May 1996 & 1800 & \phn$32^\circ$ & G1V & 1.51/1.47 & 1.0 \\*

~& [Si VI]: & & 1800 & & G4V & 1.72/1.68 & 1.0 \\*

~& H$_2+$Br$\gamma$: & & 1800 & & & 1.59/1.62 & 1.0 \\

IRAS 14352$-$1954 & Pa$\alpha$: & 6 May 1996 & 1800 & \phn\phn$0^\circ$ & G3V & 1.70/1.73 & 1.2 \\*

~& [Si VI]: & & 1800 & & G0V & 1.71/1.79 & 0.8 \\*

~& H$_2+$Br$\gamma$: & & 1800 & & G3V & 1.68/1.72 & 0.8 \\

IRAS 14394$+$5332 & Pa$\alpha$: & 5 May 1996 & 1800 & \phn$56^\circ$ & G1V & 1.34/1.30 & 0.8 \\*

~& [Si VI]: & 6 May 1996 & 1800 & & & 1.51/1.50 & 0.8 \\*

~& H$_2+$Br$\gamma$: & & 1800 & & & 1.36/1.32 & 0.7 \\

IRAS 15245$+$1019 & Pa$\alpha$: & 22 May 1997 & 1800 & $113^\circ$ & G8V & 1.11/1.12 & 1.2 \\*

~& [Si VI]: & & 1800 & & & 1.09/1.11 & 1.0 \\*

~& H$_2+$Br$\gamma$: & & 1800 & & & 1.10/1.10 & 1.0 \\

IRAS 15250$+$3609 & Pa$\alpha$: & 21 May 1997 & 1800 & $135^\circ$ & G2V & 1.03/1.01 & 1.0 \\*

~& [Si VI]: & & 2400 & & & 1.01/1.02 & 1.0 \\*

~& H$_2+$Br$\gamma$: & & 1800 & & & 1.00/1.01 & 0.7 \\

IRAS 15462$-$0450 & Pa$\alpha$: & 7 Apr 1996 & 1800 & \phn$60^\circ$ & G2.5V & 1.32/1.35 & 1.0 \\*

~& [Si VI]: & 8 Apr 1996 & 1800 & & & 1.41/1.36 & 0.8 \\*

~& H$_2+$Br$\gamma$: & 7 Apr 1996 & 1800 & & & 1.39/1.37 & 1.0 \\

IRAS 16487$+$5447 & Pa$\alpha$: & 21 May 1997 & 1800 & \phn$71^\circ$ & G5V & 1.08/1.11 & 0.7 \\*

~& [Si VI]: & & 1800 & & & 1.11/1.12 & 0.8 \\*

~& H$_2+$Br$\gamma$: & & 1800 & & G0Va & 1.17/1.20 & 0.7 \\

IRAS 17028$+$5817 & Pa$\alpha$: & 22 May 1997 & 1800 & \phn$95^\circ$ & G2V & 1.14/1.13 & 0.8 \\*

~& [Si VI]: & & 1800 & & & 1.18/1.14 & 1.0 \\*

~& H$_2+$Br$\gamma$: & & 1800 & & G0Va & 1.24/1.26 & 0.7 \\

IRAS 18470$+$3233 & Pa$\alpha$: & 31 Jul 1996 & 1800 & \phn$67^\circ$ & G8V & 1.05/1.00 & 0.8 \\*

~& [Si VI]: & & 1800 & & & 1.00/1.00 & 1.0 \\*

~& H$_2+$Br$\gamma$: & & 1800 & & & 1.01/1.00 & 1.0 \\

IRAS 19458$+$0944 & Pa$\alpha$: & 2 Aug 1996 & 1800 & $121^\circ$ & F5V+G0V & 1.09/1.09 & 0.7 \\*

~& [Si VI]: & & 1200 & & & 1.09/1.12 & 0.7 \\*

~& H$_2+$Br$\gamma$: & & 1800 & & & 1.09/1.09 & 0.7 \\

IRAS 20046$-$0623 & Pa$\alpha$: & 12 Aug 1997 & 1800 & \phn$76^\circ$ & G2V & 1.37/1.40 & 1.0 \\*

~& [Si VI]: & & 1200 & & G1V & 1.31/1.27 & 1.0 \\*

~& H$_2+$Br$\gamma$: & & 1200 & & & 1.31/1.27 & 1.0 \\

IRAS 20087$-$0308 & Pa$\alpha$: & 11 Aug 1997 & 1800 & \phn$90^\circ$ & G1V & 1.36/1.35 & 1.0 \\*

~& [Si VI]: & & 1800 & & G0V & 1.51/1.43 & 0.8 \\*

~& H$_2+$Br$\gamma$: & 12 Aug 1997 & 1800 & & G1V & 1.39/1.43 & 1.3 \\

IRAS 20414$-$1651 & Pa$\alpha$: & 1 Aug 1996 & 1800 & \phn$64^\circ$ & G2V & 1.56/1.56 & 1.0 \\*

~& [Si VI]: & & 1200 & & G1V & 1.71/1.70 & 0.8 \\*

~& H$_2+$Br$\gamma$: & & 1800 & & G2V & 1.62/1.56 & 0.8 \\

IRAS 21504$-$0628 & Pa$\alpha$: & 3 Aug 1996 & 1800 & $150^\circ$ & G0V & 1.30/1.33 & 0.8 \\*

~& [Si VI]: & & 1200 & & G8V & 1.54/1.62 & 2.0 \\*

~& H$_2+$Br$\gamma$: & & 1800 & & G0V & 1.34/1.33 & 0.7 \\

IRAS 22491$-$1808 & Pa$\alpha$: & 2 Aug 1996 & 1800 & $108^\circ$ & G8V & 1.62/1.61 & 1.0 \\*

~& [Si VI]: & & 1800 & & & 1.63/1.62 & 1.2 \\*

~& H$_2+$Br$\gamma$: & & 1800 & & & 1.60/1.61 & 1.0 \\

IRAS 23327$+$2913 & Pa$\alpha$: & 11 Aug 1997 & 1800 & $175^\circ$ & G1.5V & 1.00/1.01 & 1.0 \\*

~& [Si VI]: & & 1200 & & & 1.03/1.01 & 0.7 \\*

~& H$_2+$Br$\gamma$: & 10 Jul 1995 & 1200 & & G0V & 1.00/1.00 & 0.8 \\

IRAS 23365$+$3604 & Pa$\alpha$: & 26 Nov 1996 & 1800 & $173^\circ$ & G1.5V & 1.01/1.01 & 0.8 \\*

~& [Si VI]: & 28 Nov 1996 & 1800 & & G2V & 1.00/1.01 & 2.0 \\*

~& H$_2+$Br$\gamma$: & 26 Nov 1996 & 1800 & & G1.5V & 1.03/1.02 & 0.8 \\

\enddata
\end{deluxetable}

The most significant stellar absorption feature present in the 2 \( \mu  \)m
window for the atmospheric calibrator stars is the Br\( \gamma  \) line at
2.1655 \( \mu  \)m. While a template stellar spectrum may be adequate for removal
of this line, we instead interpolated over the line in the two-dimensional G
star calibration/flat-field spectrum before dividing the galaxy data by the
calibrator data. Multiplying the corrected galaxy spectrum by a Planck blackbody
function with a temperature matching that of the calibrator rectifies the spectrum
to a proper representation in flux density. To remove the effects of the weaker
absorption lines in the atmospheric calibrator, the galaxy spectrum was multiplied
by a normalized solar spectrum, smoothed and re-binned to the same resolution.
For the long wavelength grating setting covering the H\( _{2} \)~1--0 S(1)
and Br\( \gamma  \) lines, the CO band structure becomes significant when using
late-type G stars for the atmospheric calibration. When the atmospheric calibrator
was of type G4--G8, a linear combination of G3 V and K0 V template spectra from
\citet{templates} was used instead of the solar spectrum. Again, this template
was smoothed and re-binned to match the individual galaxy spectra.

\section{Results}

The chief results from this spectroscopic survey---that AGN appear to be rare
among ULIRGs---has been reported previously in \citet{twm99}. In the present
paper we focus attention on the range of observed properties rather than on
the collected sample as a whole.

\newcounter{figcontinue}
\begin{figure}
{\par\centering \includegraphics{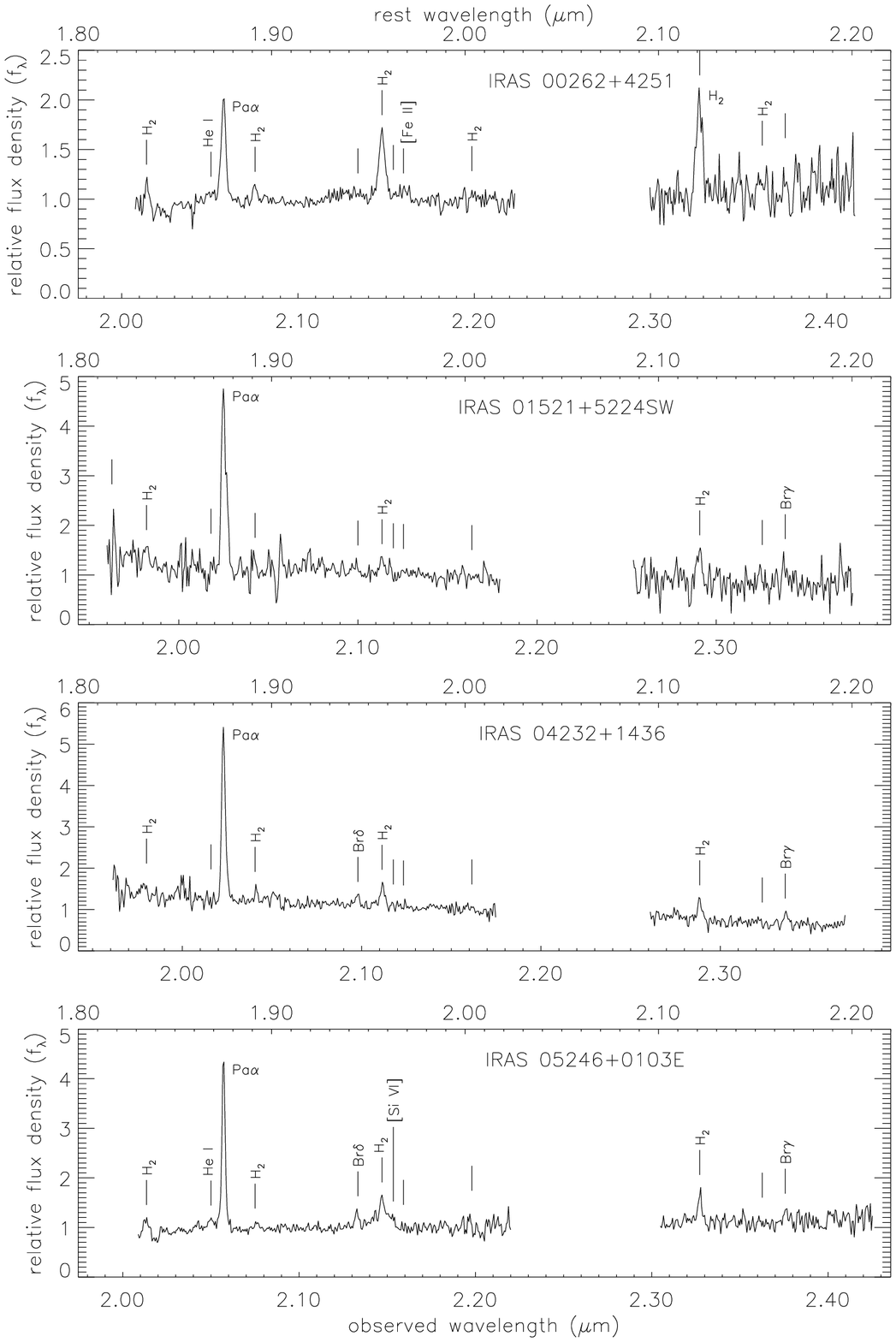} \par}
\caption{\label{fig:primary}Spectral extractions for the 33 primary ULIRG nuclei.}
\end{figure}
\setcounter{figcontinue}{\value{figure}}
\begin{figure}
\setcounter{figure}{\value{figcontinue}}
{\par\centering \includegraphics{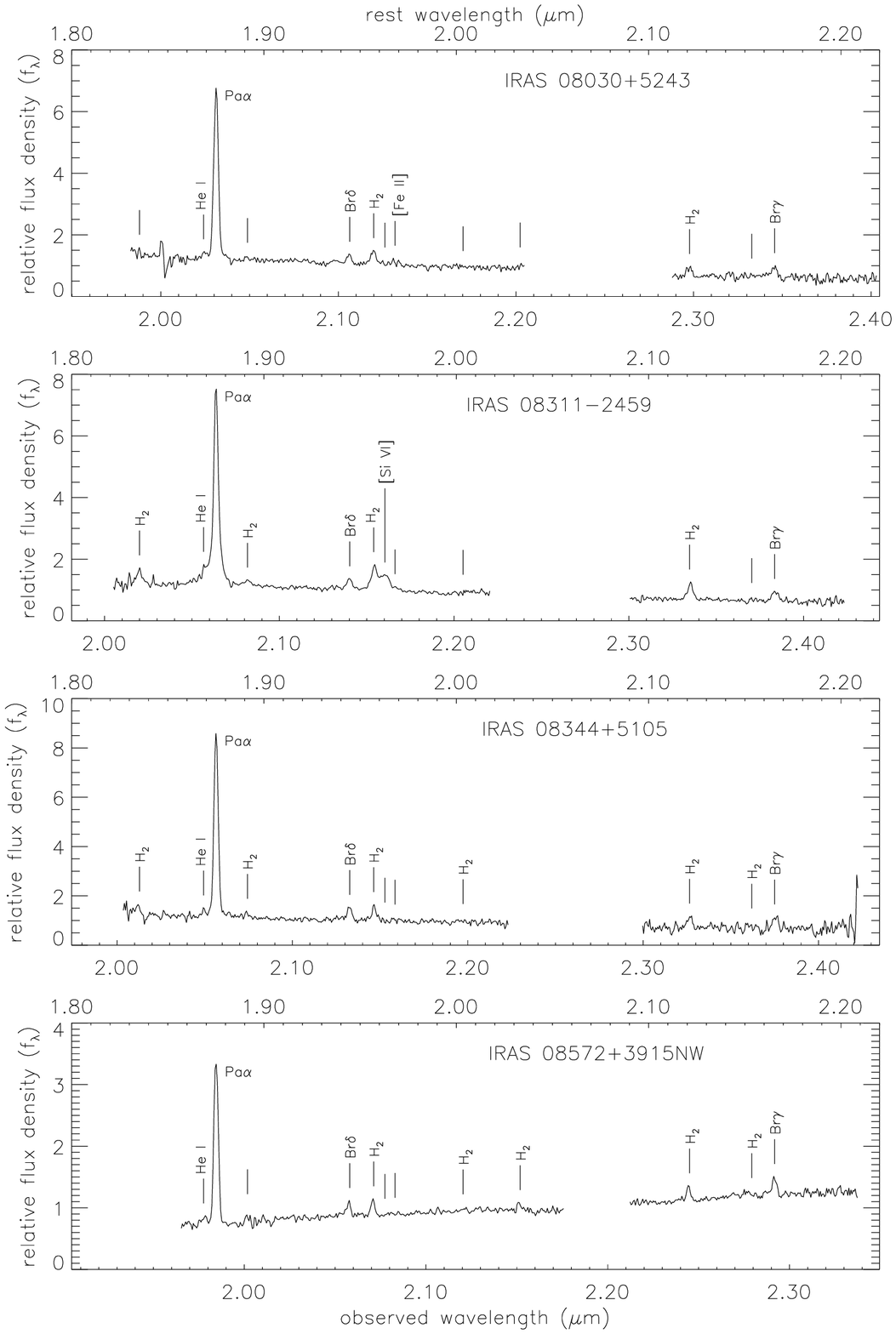} \par}
\mycaption{Figure \the\value{figure}}{continued}
\end{figure}
\begin{figure}
\setcounter{figure}{\value{figcontinue}}
{\par\centering \includegraphics{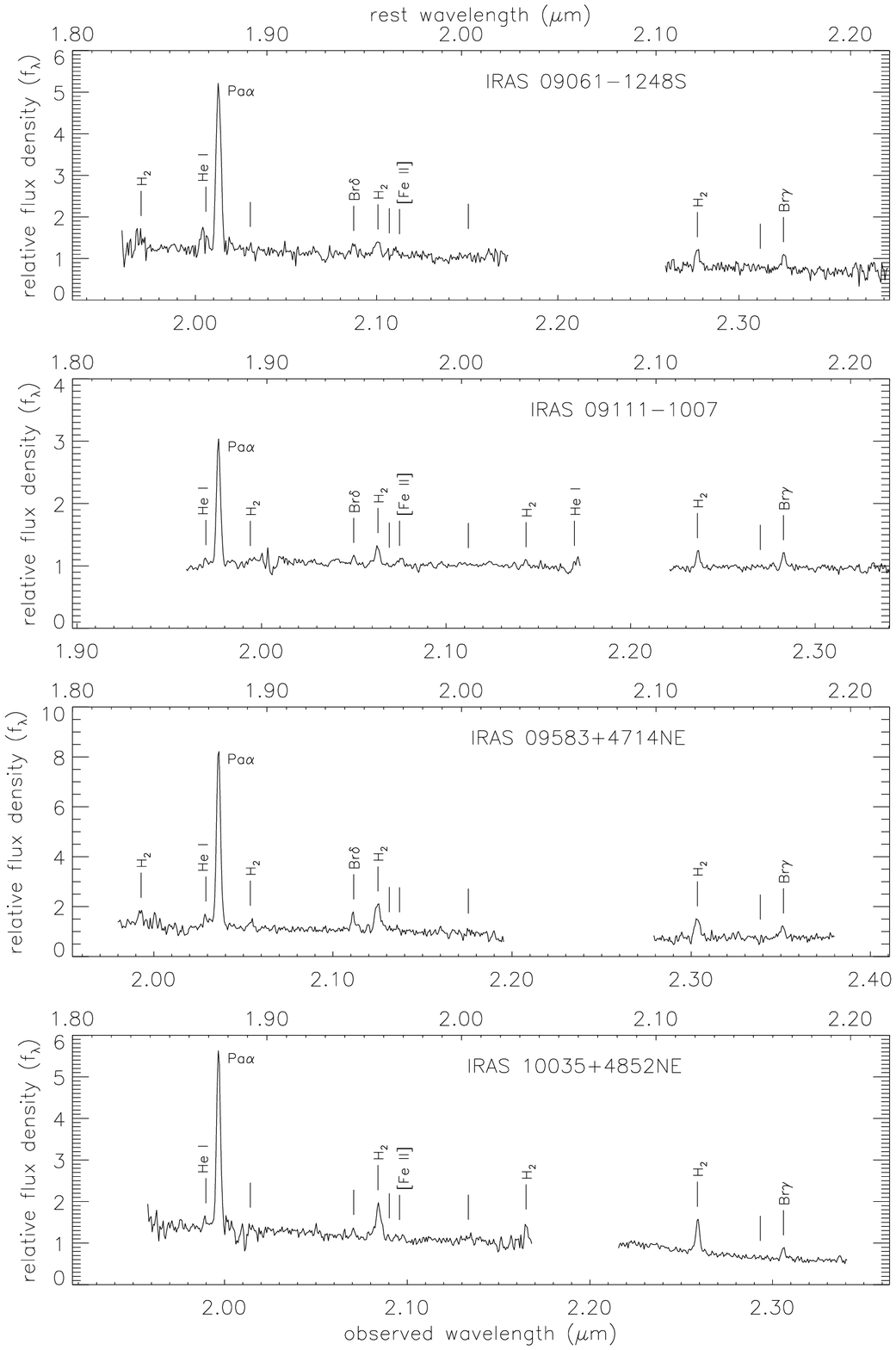} \par}
\mycaption{Figure \the\value{figure}}{continued}
\end{figure}
\begin{figure}
\setcounter{figure}{\value{figcontinue}}
{\par\centering \includegraphics{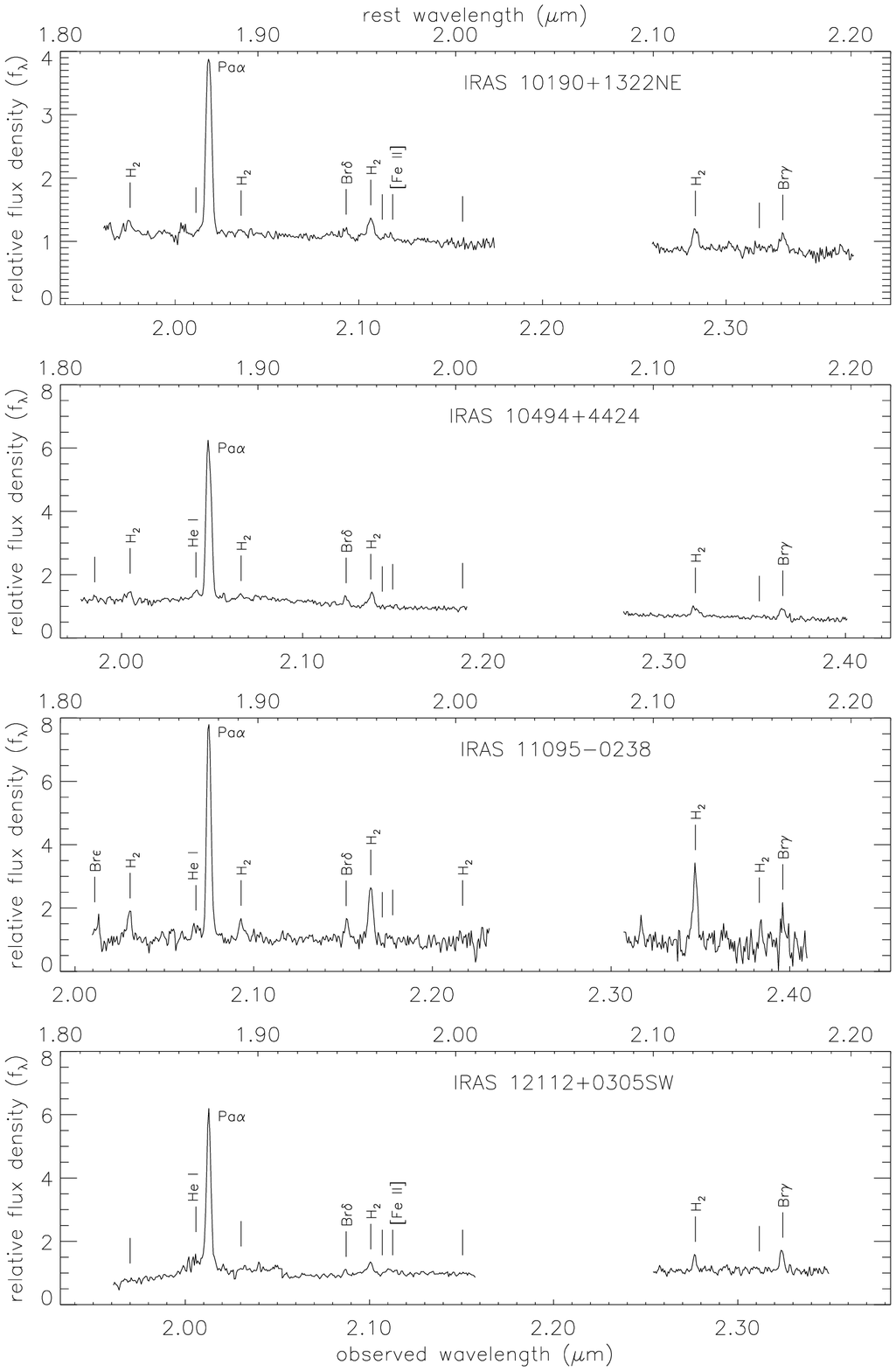} \par}
\mycaption{Figure \the\value{figure}}{continued}
\end{figure}
\begin{figure}
\setcounter{figure}{\value{figcontinue}}
{\par\centering \includegraphics{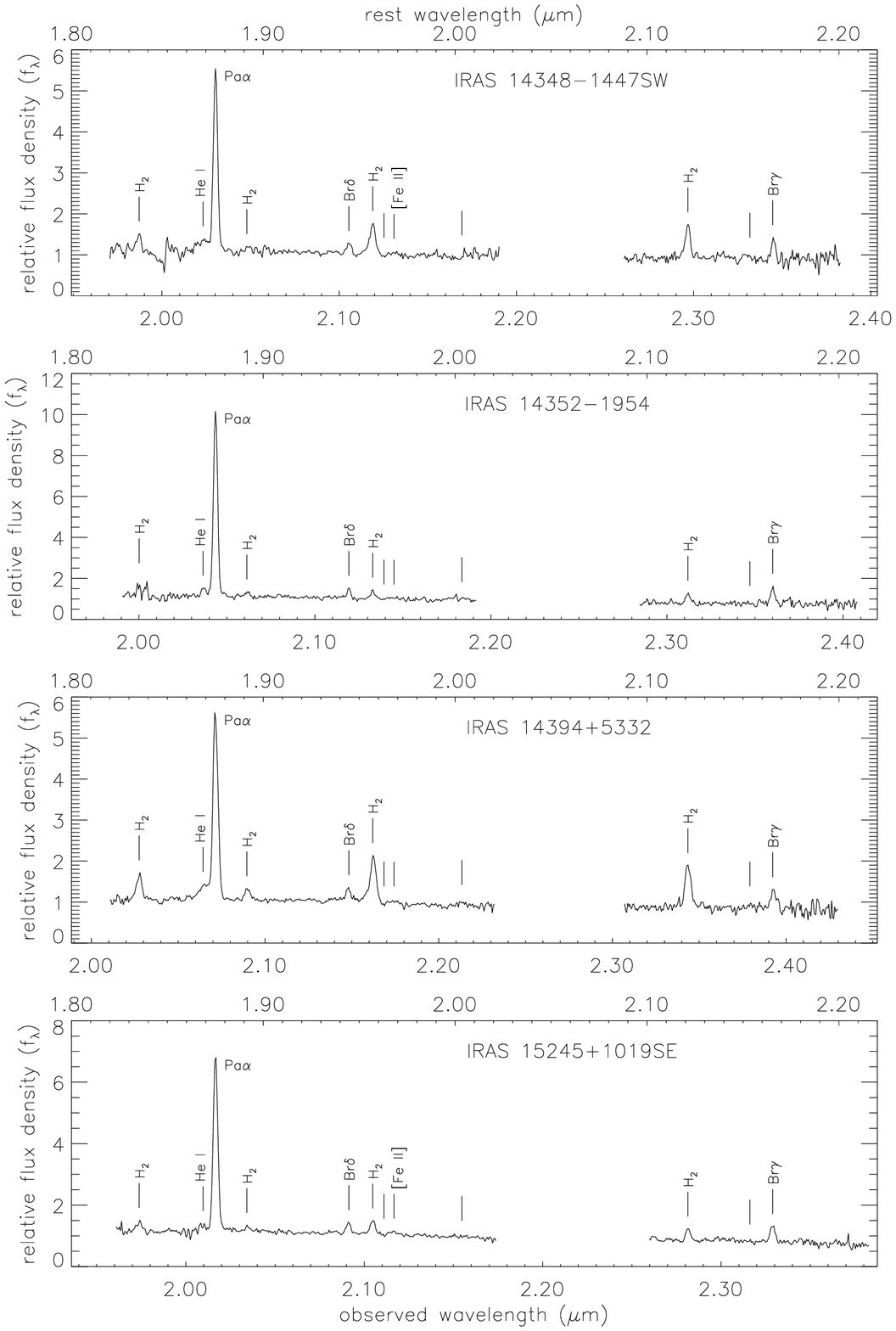} \par}
\mycaption{Figure \the\value{figure}}{continued}
\end{figure}
\begin{figure}
\setcounter{figure}{\value{figcontinue}}
{\par\centering \includegraphics{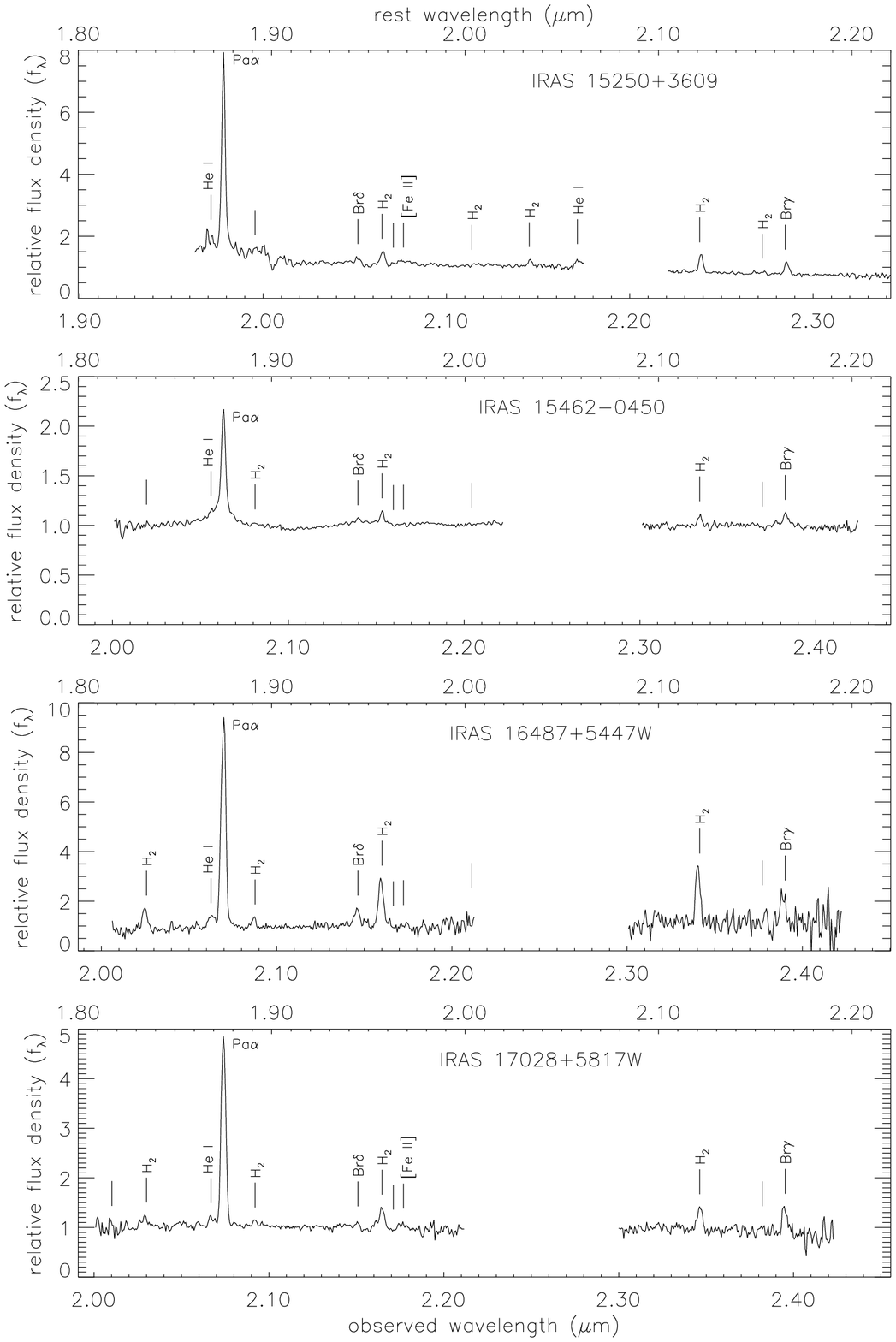} \par}
\mycaption{Figure \the\value{figure}}{continued}
\end{figure}
\begin{figure}
\setcounter{figure}{\value{figcontinue}}
{\par\centering \includegraphics{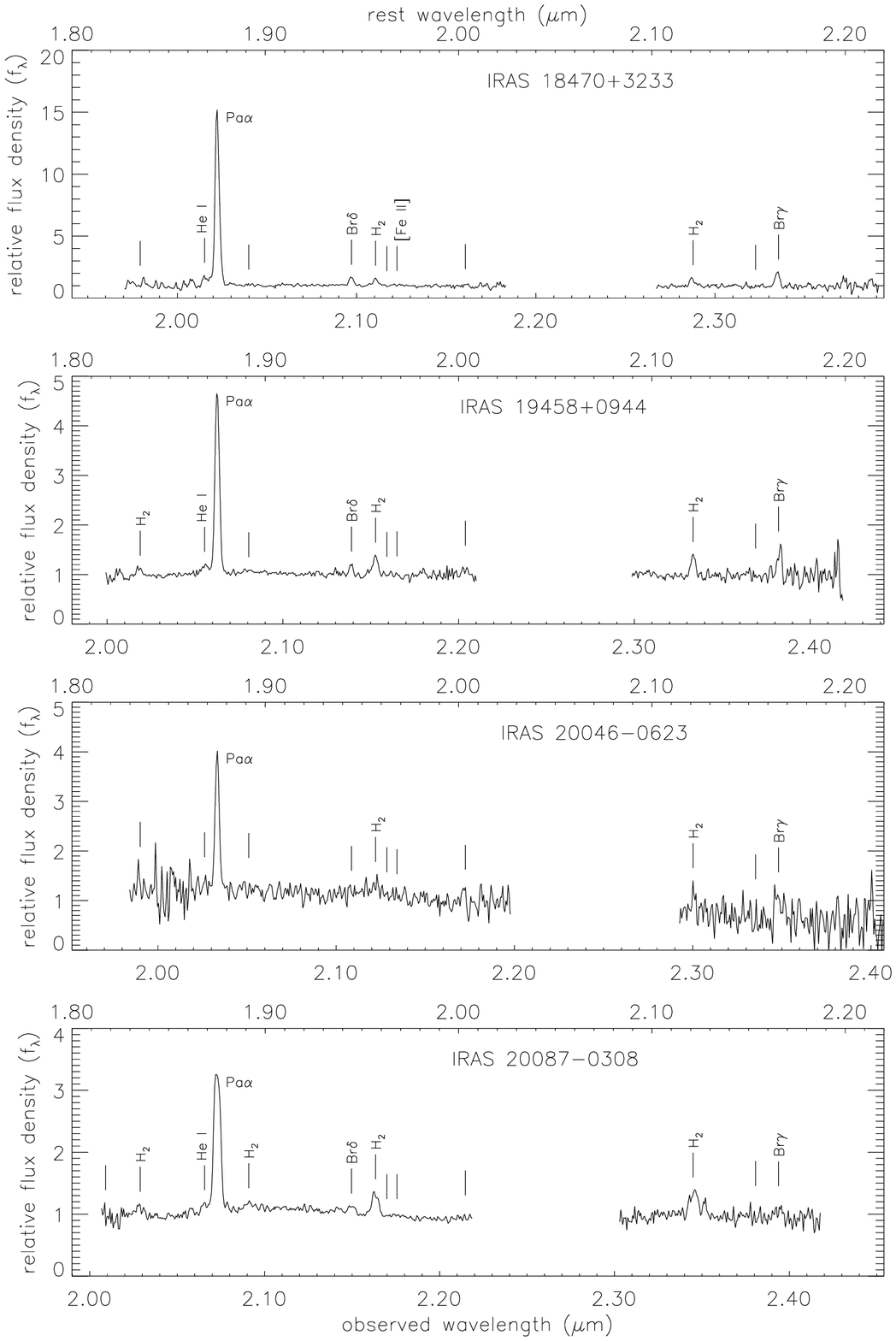} \par}
\mycaption{Figure \the\value{figure}}{continued}
\end{figure}
\begin{figure}
\setcounter{figure}{\value{figcontinue}}
{\par\centering \includegraphics{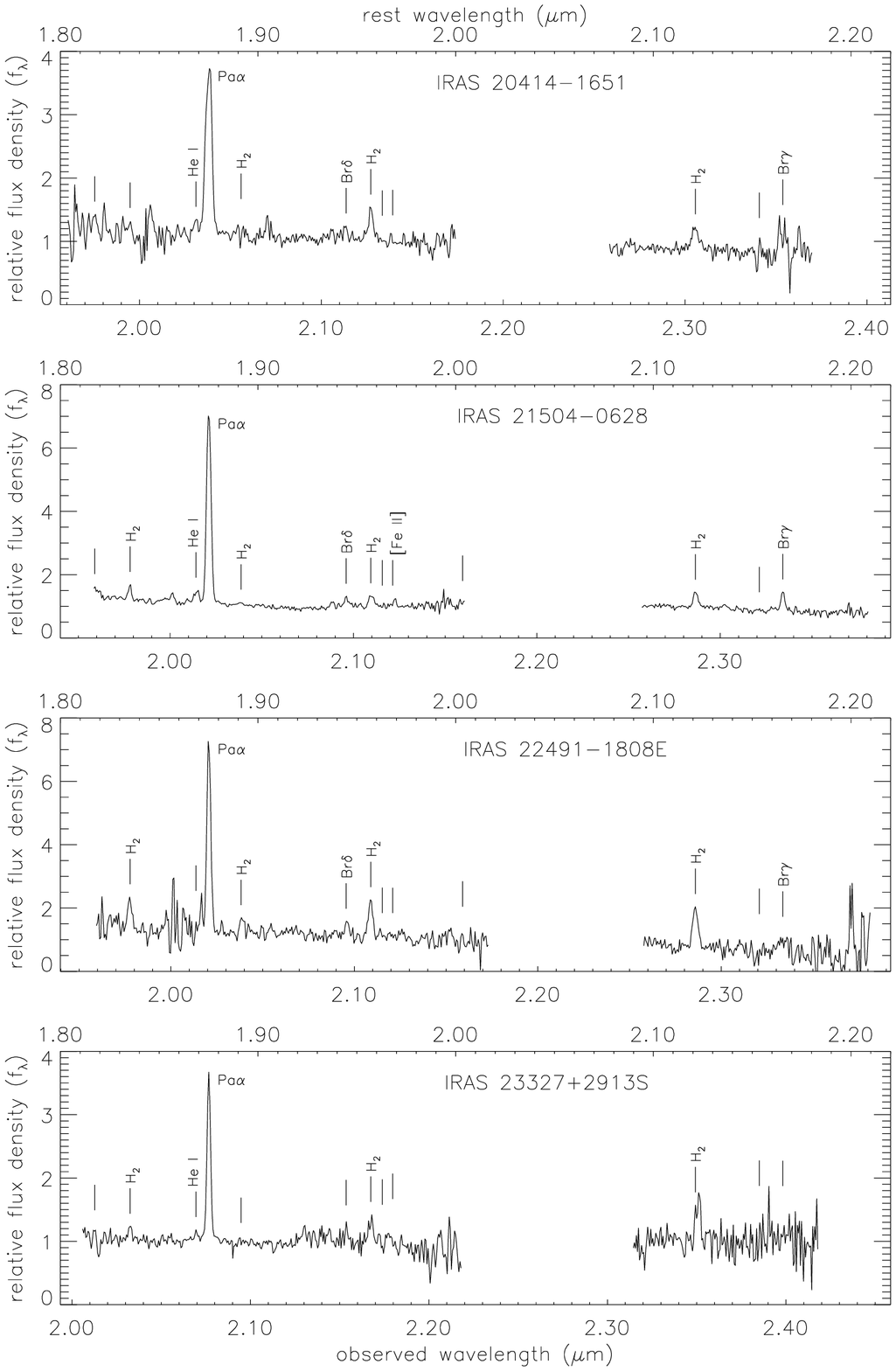} \par}
\mycaption{Figure \the\value{figure}}{continued}
\end{figure}
\begin{figure}[tbhp]
\setcounter{figure}{\value{figcontinue}}
{\par\centering \includegraphics{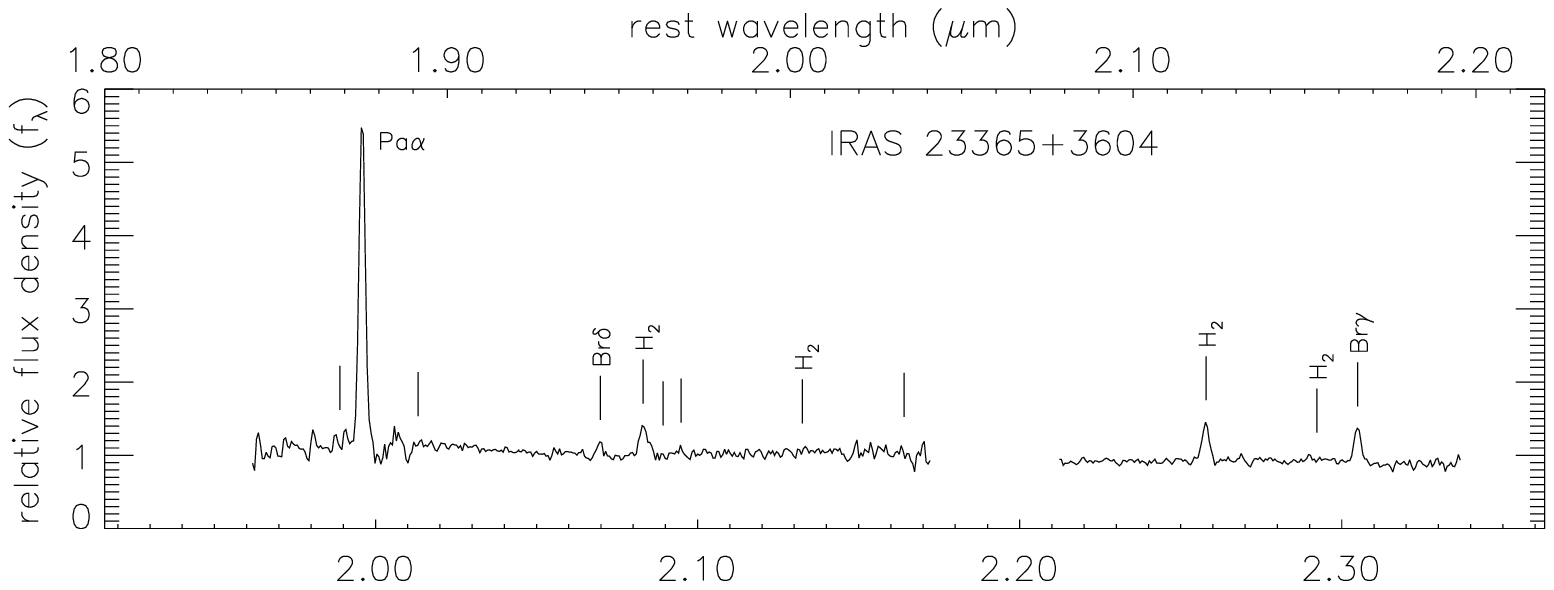} \par}
\mycaption{Figure \the\value{figure}}{continued}
\end{figure}

Figure~\ref{fig:primary} presents spectral extractions for the primary nuclei
in the 33 sample galaxies, with aperture sizes indicated in Table~\ref{tab:obs}.
Each plot is fixed in the rest frame, with the observed wavelength scale indicated
at bottom. Annotations mark the locations of every common emission line, though
in any individual spectrum only the lines with reasonable detections are labeled.
The location of the tick marks is based on the observed redshift of the Pa\( \alpha  \)
line, as is the rest wavelength scale at top. Each spectrum is composed of three
independent spectral datasets. For each of the three spectral ranges, a best
linear fit to the line-free continuum is assessed for the purpose of determining
an overall average slope for the spectrum. The individual spectra are then scaled
to lie on a common linear ``backbone'' defined by the average slope, and combined
in a noise-weighted manner across the overlap region around 1.92 \( \mu  \)m.
The vertical scale is normalized such that the continuum is at one unit of flux
density at 2.155 \( \mu  \)m, corresponding to the middle of the \( K_{s} \)
bandpass.

\begin{figure}
{\par\centering \includegraphics{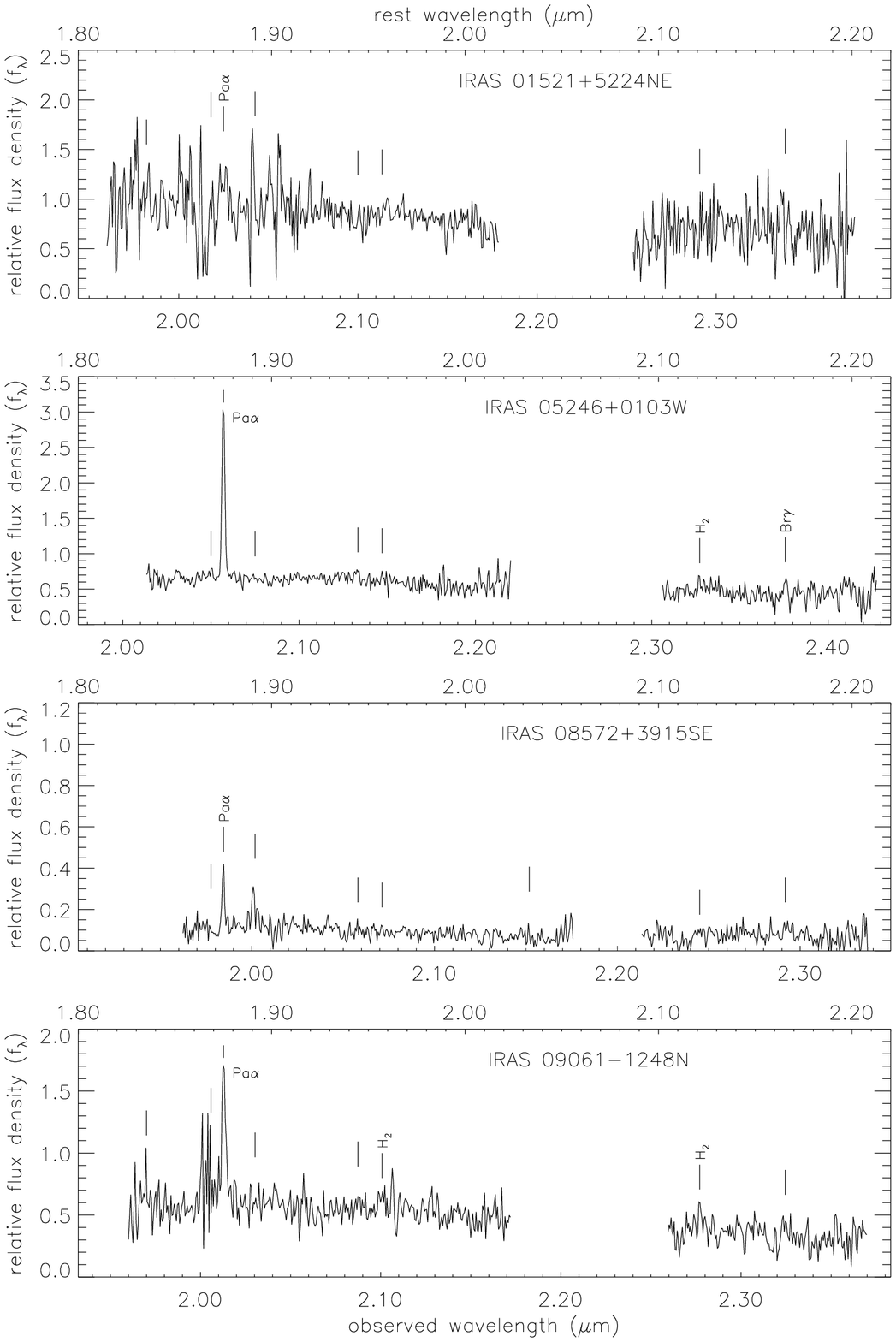} \par}
\caption{\label{fig:secondary}Spectral extractions for the 14 secondary ULIRG nuclei.}
\end{figure}
\setcounter{figcontinue}{\value{figure}}
\begin{figure}
\setcounter{figure}{\value{figcontinue}}
{\par\centering \includegraphics{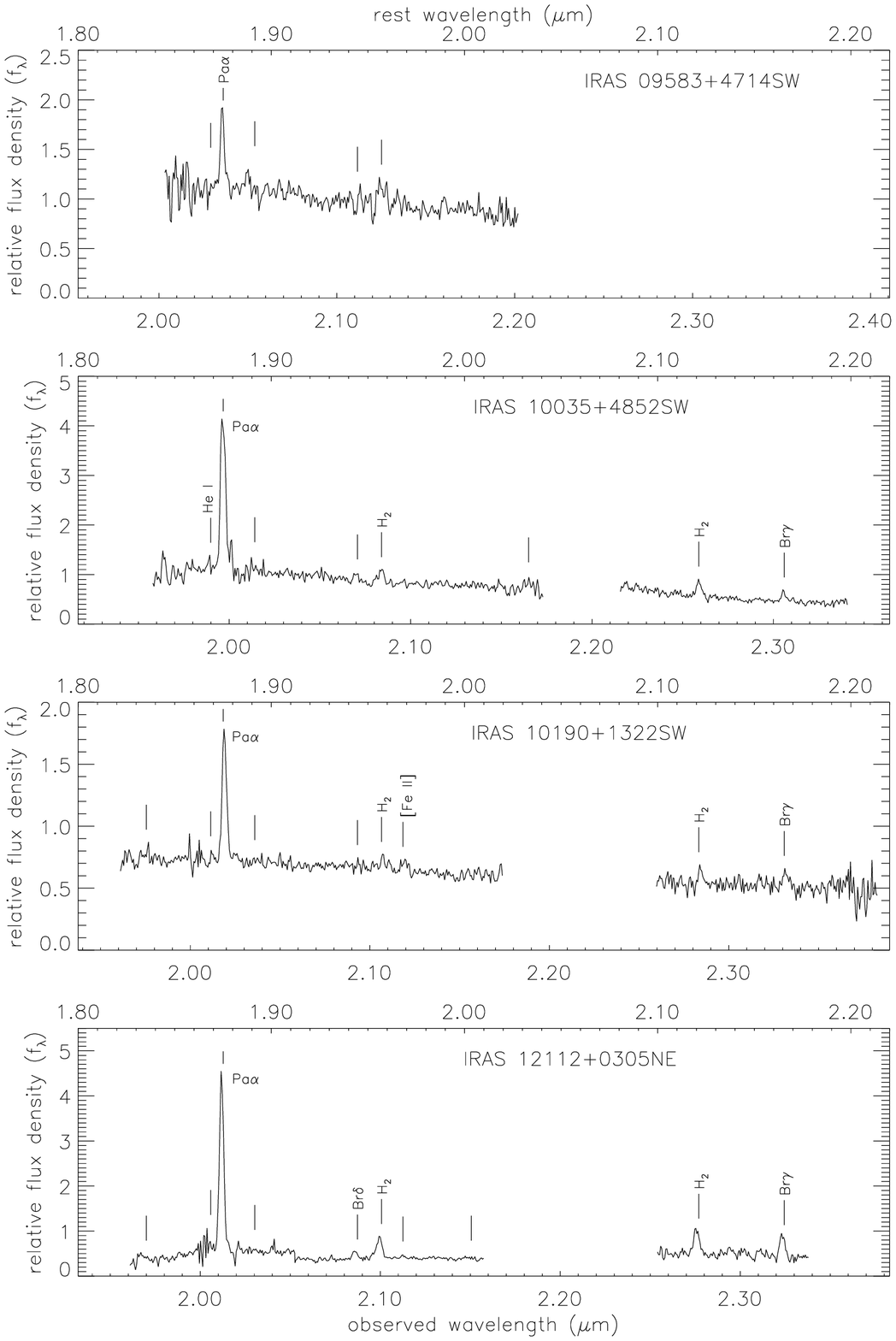} \par}
\mycaption{Figure \the\value{figure}}{continued}
\end{figure}
\begin{figure}
\setcounter{figure}{\value{figcontinue}}
{\par\centering \includegraphics{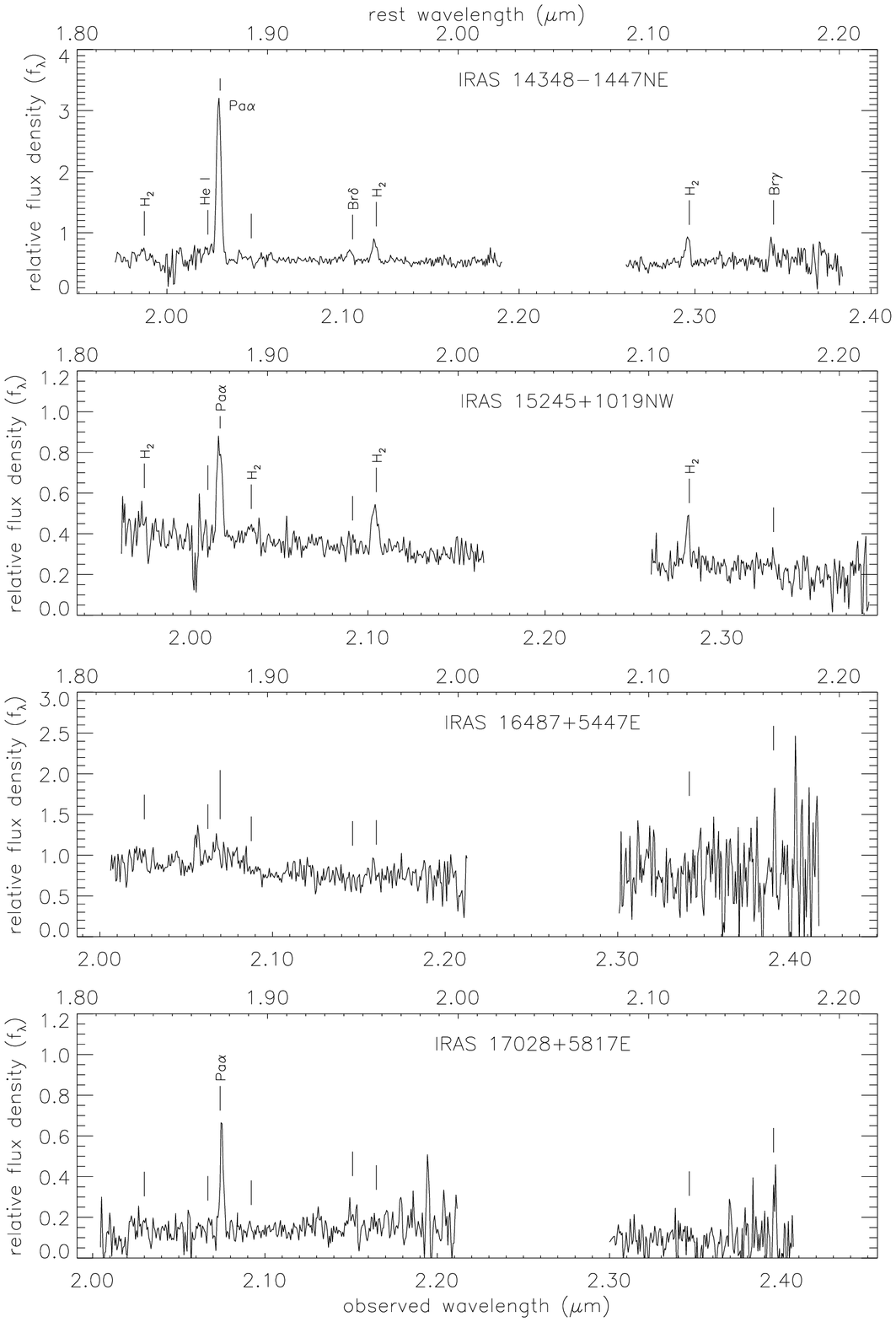} \par}
\mycaption{Figure \the\value{figure}}{continued}
\end{figure}
\begin{figure}[tbhp]
\setcounter{figure}{\value{figcontinue}}
{\par\centering \includegraphics{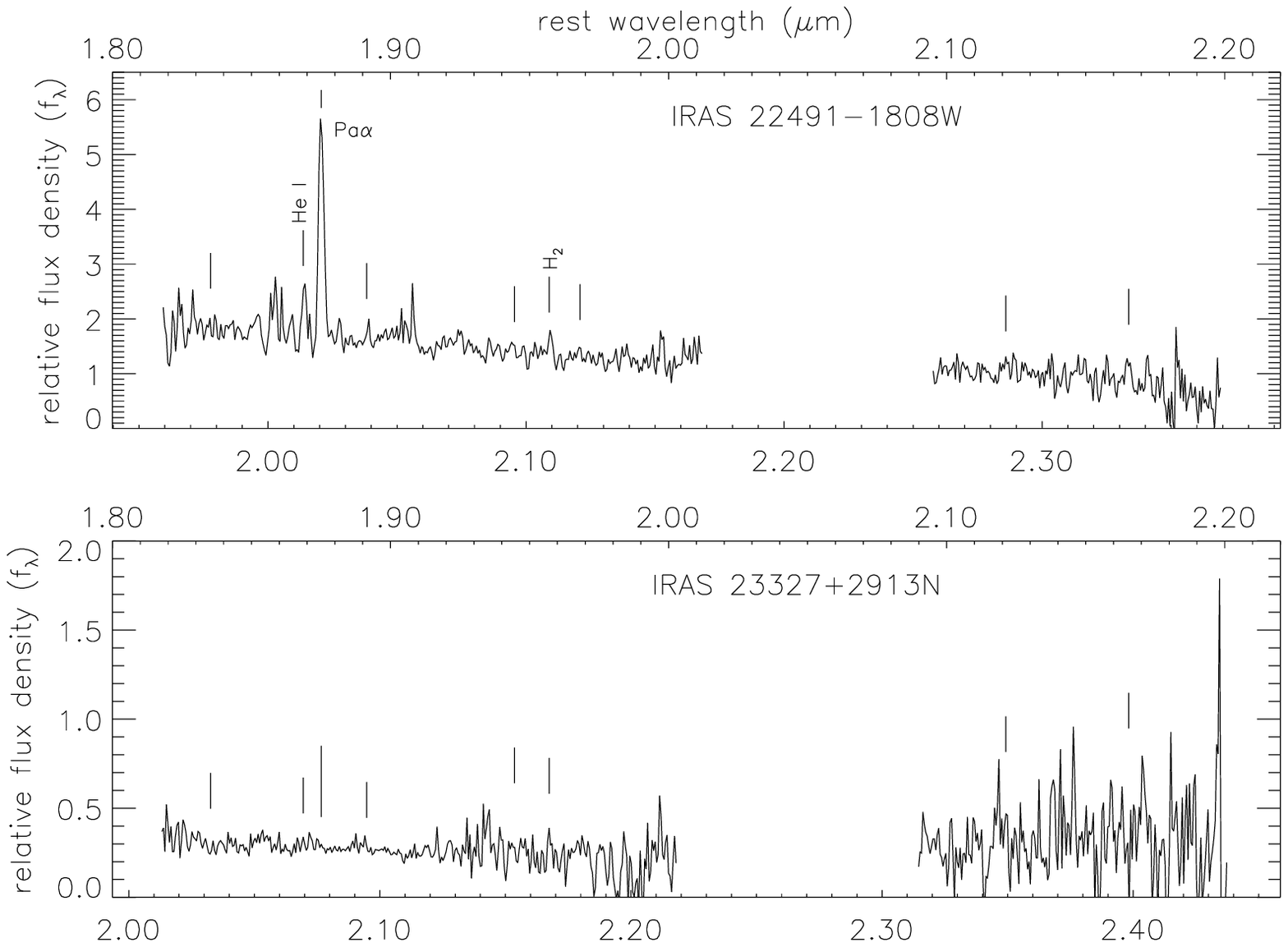} \par}
\mycaption{Figure \the\value{figure}}{continued}
\end{figure}

Figure~\ref{fig:secondary} contains spectra of the secondary nuclei in the
14 sample galaxies with a nearby galactic companion, employing the same wavelength
scale used for Figure~\ref{fig:primary}. Annotations and scalings are the
same as for Figure~\ref{fig:primary}, though some of the weaker lines are
not marked by default in this set. The tick marks denoting line positions are
based on the redshift of the Pa\( \alpha  \) line in the \emph{primary} nucleus,
so that velocity displacements between the two nuclei are more easily seen.
The numerical flux density scale for each is identical to that for the corresponding
spectrum in Figure~\ref{fig:primary}, allowing direct comparison of the relative
levels for each. The relative scaling of the three individual spectra determined
for Figure~\ref{fig:primary} are used in Figure~\ref{fig:secondary} as well.

The choice of primary versus secondary nucleus is based solely on the continuum
level, with the brighter nucleus designated as the primary. The only exception
is IRAS~22491\( - \)1808, for which the primary nucleus is identified with
the dominant line-emitting, yet weaker continuum nucleus to the east. The western
nucleus is 30\% brighter, but lacks the strong H\( _{2} \) emission that is
associated with \emph{every} other primary ULIRG nucleus. Comparison of Figures~\ref{fig:primary}
and \ref{fig:secondary} indicate that the primary nuclei, besides being the
brightest continuum sources, also exhibit apparently stronger line emission.
Yet despite the weaker line strengths, the secondary nuclei often do show appreciable
line emission activity, appearing relatively similar to the primary nuclei.
If the observed Pa\( \alpha  \) line strength is proportional to star formation
rates---and therefore total luminosity---then even though the dominant luminosity
can be associated with one nucleus, the lesser nucleus still contributes a substantial
fraction of the total luminosity. It must be realized, however, that the high
levels of extinction present in ULIRGs could easily lead one to misjudge the
identity of the primary source of infrared luminosity based simply on apparent
line strengths.

Table~\ref{tab:mofoprim} presents the line measurements from the primary nuclear
data shown in Figure~\ref{fig:primary} and Table~\ref{tab:mofosec} presents
the same data for the secondary nuclear spectra of Figure~\ref{fig:secondary}.
For each galaxy, the continuum slope and measured Pa\( \alpha  \) FWHM are
given, as well as equivalent width, FWHM, and velocity centroid of each significant
line. In this context, weaker hydrogen recombination and molecular hydrogen
vibration-rotation lines are excluded from the table as their properties follow
from those of the stronger lines. The continuum slope, \( \alpha  \), is reported
as the power-law index, where \( f_{\lambda }(\lambda )=C\lambda ^{\alpha } \).
The equivalent linear slope, with \( f_{\lambda }(\lambda )=f_{\lambda }(\lambda _{0})\left[ 1+\sigma (\lambda -\lambda _{0})\right]  \),
where \( \lambda _{0}=2.155\, \mu  \)m, is \( \sigma =\alpha /\lambda _{0} \).
The measured line properties reported in Tables~\ref{tab:mofoprim} and \ref{tab:mofosec}
are obtained by the procedures detailed in Appendix~\ref{procedures}. 

\begin{deluxetable}{lcclccccccc}
\tabletypesize{\tiny}
\rotate
\tablewidth{0pt}
\tablecaption{Measured Line Properties for Primary Nuclei\label{tab:mofoprim}}
\tablehead{\colhead{Galaxy} & \colhead{Continuum} & \colhead{Pa$\alpha$ FWHM\tablenotemark{b}} & \colhead{Rows\tablenotemark{c}} & \colhead{He I} & \colhead{Pa$\alpha$} & \colhead{H$_2$ S(3)} & \colhead{$[$Si VI$]$} & \colhead{$[$Fe II$]$} & \colhead{H$_2$ S(1)} & \colhead{Br$\gamma$} \\
& \colhead{Slope\tablenotemark{a}} & \colhead{(km s$^{-1}$)} & & \colhead{1.8689} & \colhead{1.8751} & \colhead{1.9570} & \colhead{1.9629} & \colhead{1.9670} & \colhead{2.1213} & \colhead{2.1655} \\ }
\startdata \\
IRAS 00262$+$4251 & $0.76$ & 331 & & & & & & & & \\*

& & & EW & $0.42\pm$0.1 & $3.52\pm$0.15 & $2.95\pm$0.14 & \nodata & $0.404\pm$0.088 & $3.88\pm$0.24 & \nodata \\*

& & & FWHM & 420$\pm$240 & 366$\pm$27 & 480$\pm$42 & & 260$\pm$160 & 380$\pm$56 & \\*

& & & $cz$ & $-67\pm$78 & $29204\pm$18 & $-18\pm$29 & & $187\pm$63 & $-44\pm$27 & \\

IRAS 01521$+$5224S & $-1.94$ & 394 & & & & & & & & \\*

& & & EW & \nodata & $12.02\pm$0.44 & $0.79\pm$0.14 & \nodata & \nodata & $2.44\pm$0.36 & $1.38\pm$0.49 \\*

& & & FWHM & & 356$\pm$23 & 170$\pm$96 & & & 290$\pm$120 & 0$\pm$110 \\*

& & & $cz$ & & $24011\pm$16 & $-58\pm$46 & & & $-101\pm$34 & $-95\pm$76 \\

IRAS 04232$+$1436 & $-4.07$ & 183 & & & & & & & & \\*

& & & EW & \nodata & $8.74\pm$0.25 & $1.34\pm$0.15 & \nodata & \nodata & $1.76\pm$0.16 & $1.4\pm$0.26 \\*

& & & FWHM & & 218$\pm$18 & 195$\pm$70 & & & 203$\pm$51 & 220$\pm$100 \\*

& & & $cz$ & & $23666\pm$15 & $25\pm$33 & & & $-11\pm$26 & $63\pm$46 \\

IRAS 05246$+$0103E & $1.18$ & 110 & & & & & & & & \\*

& & & EW & $1.09\pm$0.13 & $8.89\pm$0.19 & $3.15\pm$0.26 & $0.86\pm$0.2 & \nodata & $1.53\pm$0.14 & $0.61\pm$0.24 \\*

& & & FWHM & 497$\pm$78 & 190.1$\pm$9.6 & 575$\pm$77 & 310$\pm$260 & & 161$\pm$40 & 210$\pm$210 \\*

& & & $cz$ & $-83\pm$51 & $29105\pm$22 & $255\pm$47 & $-180\pm$55 & & $27\pm$38 & $70\pm$110 \\

IRAS 08030$+$5243 & $-4.2$ & 322 & & & & & & & & \\*

& & & EW & $0.53\pm$0.13 & $14.46\pm$0.16 & $1.3\pm$0.11 & \nodata & $0.39\pm$0.1 & $1.36\pm$0.22 & $2.32\pm$0.28 \\*

& & & FWHM & 340$\pm$160 & 326.2$\pm$6.8 & 331$\pm$77 & & 0$\pm$91 & 270$\pm$130 & 380$\pm$110 \\*

& & & $cz$ & $97\pm$71 & $24959\pm$12 & $-29\pm$34 & & $196\pm$87 & $-47\pm$31 & $-173\pm$44 \\

IRAS 08311$-$2459 & $-3.46$ & 314 & & & & & & & & \\*

& & & EW & \nodata & $22.69\pm$0.29 & $3.68\pm$0.15 & $3.005\pm$0.088 & $0.488\pm$0.078 & $2.734\pm$0.092 & $2.18\pm$0.25 \\*

& & & FWHM & & 477$\pm$10 & 501$\pm$41 & 708$\pm$52 & 360$\pm$130 & 351$\pm$27 & 470$\pm$170 \\*

& & & $cz$ & & $30114\pm$16 & $129\pm$26 & $78\pm$20 & $337\pm$45 & $104\pm$21 & $94\pm$34 \\

IRAS 08344$+$5105 & $-3.13$ & 287 & & & & & & & & \\*

& & & EW & $0.67\pm$0.13 & $20.08\pm$0.26 & $1.64\pm$0.13 & \nodata & \nodata & $2.16\pm$0.27 & $2.32\pm$0.36 \\*

& & & FWHM & 60$\pm$110 & 300$\pm$13 & 196$\pm$40 & & & 310$\pm$100 & 310$\pm$140 \\*

& & & $cz$ & $-45\pm$39 & $29002\pm$12 & $60\pm$28 & & & $0\pm$30 & $52\pm$35 \\

IRAS 08572$+$3915N & $3.26$ & 340 & & & & & & & & \\*

& & & EW & $0.84\pm$0.13 & $11.192\pm$0.1 & $0.695\pm$0.062 & \nodata & \nodata & $0.448\pm$0.027 & $0.762\pm$0.072 \\*

& & & FWHM & 440$\pm$110 & 334.8$\pm$8 & 139$\pm$45 & & & 135$\pm$30 & 271$\pm$62 \\*

& & & $cz$ & $-51\pm$53 & $17493\pm$19 & $-69\pm$37 & & & $-78\pm$31 & $-56\pm$37 \\

IRAS 09061$-$1248S & $-3.28$ & 371 & & & & & & & & \\*

& & & EW & $1.16\pm$0.5 & $10.33\pm$0.25 & $1.215\pm$0.092 & \nodata & $0.56\pm$0.19 & $1.53\pm$0.13 & $1.47\pm$0.23 \\*

& & & FWHM & 240$\pm$280 & 320$\pm$24 & 421$\pm$99 & & 100$\pm$240 & 266$\pm$58 & 260$\pm$110 \\*

& & & $cz$ & $-260\pm$120 & $22017\pm$13 & $-81\pm$26 & & $-72\pm$96 & $34\pm$23 & $7\pm$45 \\

IRAS 09111$-$1007 & $-0.44$ & 223 & & & & & & & & \\*

& & & EW & $0.42\pm$0.11 & $5.53\pm$0.1 & $0.736\pm$0.033 & \nodata & $0.295\pm$0.052 & $0.682\pm$0.035 & $0.59\pm$0.062 \\*

& & & FWHM & 290$\pm$210 & 246$\pm$13 & 189$\pm$27 & & 310$\pm$91 & 164$\pm$25 & 168$\pm$66 \\*

& & & $cz$ & $67\pm$99 & $16231\pm$12 & $-12\pm$17 & & $244\pm$51 & $78\pm$20 & $54\pm$26 \\

IRAS 09583$+$4714NE & $-3.16$ & 247 & & & & & & & & \\*

& & & EW & $0.74\pm$0.11 & $16.75\pm$0.2 & $3.58\pm$0.22 & \nodata & \nodata & $4.06\pm$0.25 & $1.85\pm$0.25 \\*

& & & FWHM & 0$\pm$63 & 263.6$\pm$6.1 & 411$\pm$74 & & & 384$\pm$59 & 256$\pm$98 \\*

& & & $cz$ & $-78\pm$24 & $25754\pm$12 & $25\pm$25 & & & $13\pm$28 & $-134\pm$38 \\

IRAS 10035$+$4852NE & $-4.78$ & 237 & & & & & & & & \\*

& & & EW & $0.454\pm$0.049 & $8.71\pm$0.19 & $2.46\pm$0.13 & \nodata & $0.21\pm$0.11 & $2.915\pm$0.089 & $1.08\pm$0.15 \\*

& & & FWHM & 0$\pm$61 & 254$\pm$12 & 315$\pm$43 & & 0$\pm$240 & 256$\pm$18 & 201$\pm$71 \\*

& & & $cz$ & $-43\pm$29 & $19435\pm$17 & $38\pm$35 & & $180\pm$170 & $-2\pm$28 & $46\pm$44 \\

IRAS 10190$+$1322E & $-1.83$ & 445 & & & & & & & & \\*

& & & EW & \nodata & $9.6\pm$0.13 & $1.164\pm$0.081 & \nodata & $0.154\pm$0.072 & $1.248\pm$0.096 & $1.12\pm$0.11 \\*

& & & FWHM & & 437$\pm$16 & 412$\pm$72 & & 0$\pm$130 & 336$\pm$65 & 320$\pm$91 \\*

& & & $cz$ & & $22889\pm$19 & $12\pm$37 & & $-50\pm$140 & $7\pm$32 & $22\pm$34 \\

IRAS 10494$+$4424 & $-3.98$ & 430 & & & & & & & & \\*

& & & EW & $1.04\pm$0.11 & $14.96\pm$0.12 & $1.612\pm$0.072 & \nodata & \nodata & $1.89\pm$0.11 & $1.73\pm$0.16 \\*

& & & FWHM & 510$\pm$160 & 404.2$\pm$7.1 & 377$\pm$38 & & & 427$\pm$65 & 364$\pm$75 \\*

& & & $cz$ & $-51\pm$34 & $27675\pm$16 & $-2\pm$29 & & & $-3\pm$29 & $-51\pm$31 \\

IRAS 11095$-$0238 & $-1.2$ & 269 & & & & & & & & \\*

& & & EW & $1.62\pm$0.3 & $20.38\pm$0.39 & $5.65\pm$0.25 & \nodata & \nodata & $8.56\pm$0.61 & $2.5\pm$1 \\*

& & & FWHM & 360$\pm$180 & 286$\pm$12 & 350$\pm$49 & & & 241$\pm$43 & 0$\pm$100 \\*

& & & $cz$ & $-38\pm$64 & $31968\pm$18 & $-59\pm$26 & & & $-73\pm$35 & $80\pm$130 \\

IRAS 12112$+$0305SW & $1.17$ & 159 & & & & & & & & \\*

& & & EW & $1.68\pm$0.61 & $13.97\pm$0.39 & $1.095\pm$0.086 & \nodata & $0.34\pm$0.14 & $0.957\pm$0.071 & $1.57\pm$0.15 \\*

& & & FWHM & 380$\pm$340 & 303$\pm$21 & 295$\pm$63 & & 220$\pm$290 & 0$\pm$31 & 185$\pm$53 \\*

& & & $cz$ & $80\pm$140 & $21980\pm$21 & $-65\pm$42 & & $-100\pm$190 & $-19\pm$30 & $-51\pm$33 \\

IRAS 14348$-$1447SW & $-0.98$ & 227 & & & & & & & & \\*

& & & EW & $1.47\pm$0.15 & $12.78\pm$0.19 & $3.15\pm$0.12 & \nodata & $0.244\pm$0.083 & $3.11\pm$0.14 & $1.35\pm$0.2 \\*

& & & FWHM & 550$\pm$120 & 236.8$\pm$8.4 & 450$\pm$35 & & 120$\pm$210 & 334$\pm$36 & 127$\pm$80 \\*

& & & $cz$ & $150\pm$43 & $24802\pm$22 & $-65\pm$35 & & $170\pm$110 & $-46\pm$30 & $35\pm$40 \\

IRAS 14352$-$1954 & $-2.49$ & 192 & & & & & & & & \\*

& & & EW & $1.102\pm$0.089 & $23.54\pm$0.23 & $1.19\pm$0.11 & \nodata & \nodata & $1.88\pm$0.15 & $2.97\pm$0.26 \\*

& & & FWHM & 258$\pm$55 & 205.3$\pm$4.9 & 193$\pm$58 & & & 243$\pm$48 & 166$\pm$46 \\*

& & & $cz$ & $19\pm$33 & $26942\pm$17 & $9\pm$37 & & & $35\pm$34 & $6\pm$32 \\

IRAS 14394$+$5332 & $-1.41$ & 363 & & & & & & & & \\*

& & & EW & $1.602\pm$0.064 & $15.75\pm$0.13 & $5.03\pm$0.097 & \nodata & $0.143\pm$0.08 & $4.9\pm$0.16 & $2.14\pm$0.27 \\*

& & & FWHM & 558$\pm$77 & 381.8$\pm$8.7 & 490$\pm$22 & & 160$\pm$280 & 426$\pm$33 & 370$\pm$110 \\*

& & & $cz$ & $-137\pm$33 & $31373\pm$21 & $-5\pm$31 & & $210\pm$170 & $61\pm$33 & $127\pm$47 \\

IRAS 15245$+$1019E & $-2.34$ & 283 & & & & & & & & \\*

& & & EW & $0.277\pm$0.069 & $15.32\pm$0.14 & $1.324\pm$0.067 & \nodata & $0.341\pm$0.078 & $1.242\pm$0.088 & $2.08\pm$0.13 \\*

& & & FWHM & 0$\pm$75 & 294.6$\pm$7.3 & 307$\pm$46 & & 300$\pm$210 & 247$\pm$50 & 304$\pm$51 \\*

& & & $cz$ & $25\pm$39 & $22634\pm$16 & $-19\pm$30 & & $137\pm$67 & $25\pm$27 & $-15\pm$25 \\

IRAS 15250$+$3609 & $-3.7$ & 57 & & & & & & & & \\*

& & & EW & $0.43\pm$0.23 & $9.49\pm$0.16 & $1.45\pm$0.14 & \nodata & $0.55\pm$0.1 & $1.928\pm$0.059 & $1.429\pm$0.077 \\*

& & & FWHM & 0$\pm$180 & 147.8$\pm$9.9 & 323$\pm$88 & & 400$\pm$180 & 223$\pm$19 & 204$\pm$34 \\*

& & & $cz$ & $-70\pm$200 & $16535\pm$17 & $6\pm$35 & & $-15\pm$60 & $58\pm$32 & $100\pm$33 \\

IRAS 15462$-$0450 & $-0.05$ & 285 & & & & & & & & \\*

& & & EW & \nodata & $5.232\pm$0.085 & $0.48\pm$0.029 & \nodata & \nodata & $0.268\pm$0.024 & $0.516\pm$0.06 \\*

& & & FWHM & & 535$\pm$15 & 297$\pm$53 & & & 102$\pm$43 & 360$\pm$110 \\*

& & & $cz$ & & $29917\pm$24 & $229\pm$40 & & & $158\pm$32 & $237\pm$44 \\

IRAS 16487$+$5447W & $1.47$ & 347 & & & & & & & & \\*

& & & EW & $1.96\pm$0.18 & $29.02\pm$0.34 & $7.13\pm$0.29 & \nodata & \nodata & $5.95\pm$0.39 & $3.46\pm$0.89 \\*

& & & FWHM & 500$\pm$140 & 332.3$\pm$9.1 & 374$\pm$36 & & & 265$\pm$54 & 230$\pm$140 \\*

& & & $cz$ & $29\pm$40 & $31106\pm$23 & $-52\pm$33 & & & $-134\pm$30 & $-195\pm$78 \\

IRAS 17028$+$5817W & $-0.61$ & 339 & & & & & & & & \\*

& & & EW & $0.631\pm$0.097 & $11.89\pm$0.13 & $1.26\pm$0.078 & \nodata & $0.35\pm$0.13 & $1.53\pm$0.14 & $2.13\pm$0.25 \\*

& & & FWHM & 320$\pm$130 & 315.6$\pm$8.7 & 250$\pm$48 & & 140$\pm$170 & 301$\pm$74 & 336$\pm$94 \\*

& & & $cz$ & $46\pm$46 & $31805\pm$24 & $3\pm$33 & & $120\pm$130 & $6\pm$39 & $22\pm$45 \\

IRAS 18470$+$3233 & $-0.45$ & 228 & & & & & & & & \\*

& & & EW & $2.21\pm$0.4 & $41.07\pm$0.44 & $2.04\pm$0.16 & \nodata & $0.41\pm$0.16 & $2.46\pm$0.17 & $3.52\pm$0.25 \\*

& & & FWHM & 200$\pm$120 & 258.2$\pm$6.9 & 305$\pm$68 & & 200$\pm$300 & 319$\pm$49 & 249$\pm$50 \\*

& & & $cz$ & $-9\pm$51 & $23517\pm$21 & $11\pm$44 & & $140\pm$120 & $11\pm$34 & $-101\pm$30 \\

IRAS 19458$+$0944 & $-0.25$ & 269 & & & & & & & & \\*

& & & EW & $0.668\pm$0.086 & $11.093\pm$0.075 & $1.46\pm$0.13 & \nodata & \nodata & $1.236\pm$0.089 & $1.74\pm$0.29 \\*

& & & FWHM & 390$\pm$180 & 272.9$\pm$5.1 & 324$\pm$84 & & & 237$\pm$50 & 150$\pm$110 \\*

& & & $cz$ & $-27\pm$58 & $29982\pm$35 & $-83\pm$59 & & & $3\pm$47 & $73\pm$53 \\

IRAS 20046$-$0623 & $-3.99$ & 244 & & & & & & & & \\*

& & & EW & \nodata & $6.3\pm$0.21 & $1.16\pm$0.2 & \nodata & $0.45\pm$0.2 & $1.63\pm$0.44 & $3.67\pm$1 \\*

& & & FWHM & & 247$\pm$27 & 190$\pm$100 & & 0$\pm$140 & 0$\pm$88 & 310$\pm$170 \\*

& & & $cz$ & & $25293\pm$14 & $-16\pm$43 & & $50\pm$140 & $6\pm$60 & $100\pm$80 \\

IRAS 20087$-$0308 & $-0.3$ & 595 & & & & & & & & \\*

& & & EW & $0.465\pm$0.093 & $10.22\pm$0.14 & $1.611\pm$0.067 & \nodata & \nodata & $2.11\pm$0.15 & $0.64\pm$0.24 \\*

& & & FWHM & 270$\pm$150 & 574$\pm$23 & 475$\pm$49 & & & 570$\pm$110 & 280$\pm$280 \\*

& & & $cz$ & $-92\pm$48 & $31600\pm$20 & $-41\pm$33 & & & $64\pm$30 & $36\pm$97 \\

IRAS 20414$-$1651 & $-1.93$ & 515 & & & & & & & & \\*

& & & EW & $0.82\pm$0.16 & $10.47\pm$0.21 & $1.64\pm$0.16 & \nodata & \nodata & $1.71\pm$0.17 & $2.23\pm$0.63 \\*

& & & FWHM & 400$\pm$190 & 488$\pm$20 & 287$\pm$82 & & & 460$\pm$120 & 170$\pm$180 \\*

& & & $cz$ & $53\pm$71 & $26043\pm$25 & $16\pm$49 & & & $35\pm$46 & $-137\pm$92 \\

IRAS 21504$-$0628 & $-1.29$ & 256 & & & & & & & & \\*

& & & EW & $1.07\pm$0.14 & $16.82\pm$0.14 & $0.87\pm$0.14 & \nodata & \nodata & $1.684\pm$0.099 & $1.74\pm$0.13 \\*

& & & FWHM & 230$\pm$100 & 260.8$\pm$5.6 & 250$\pm$130 & & & 310$\pm$43 & 144$\pm$40 \\*

& & & $cz$ & $-17\pm$52 & $23352\pm$26 & $46\pm$44 & & & $43\pm$43 & $-4\pm$41 \\

IRAS 22491$-$1808E & $-4.12$ & 128 & & & & & & & & \\*

& & & EW & \nodata & $13.21\pm$0.38 & $3.06\pm$0.23 & \nodata & \nodata & $5.26\pm$0.37 & $3.04\pm$0.83 \\*

& & & FWHM & & 131$\pm$18 & 0$\pm$52 & & & 302$\pm$60 & 430$\pm$350 \\*

& & & $cz$ & & $23277\pm$24 & $-35\pm$41 & & & $-2\pm$33 & $10\pm$77 \\

IRAS 23327$+$2913S & $0.08$ & 151 & & & & & & & & \\*

& & & EW & $0.43\pm$0.11 & $6.85\pm$0.12 & $0.87\pm$0.26 & \nodata & \nodata & $1.92\pm$0.36 & \nodata \\*

& & & FWHM & 170$\pm$130 & 193$\pm$9 & 110$\pm$120 & & & 210$\pm$120 & \\*

& & & $cz$ & $68\pm$76 & $32235\pm$18 & $-1\pm$77 & & & $219\pm$51 & \\

IRAS 23365$+$3604 & $-1.46$ & 186 & & & & & & & & \\*

& & & EW & \nodata & $9.284\pm$0.083 & $1.291\pm$0.079 & \nodata & \nodata & $1.493\pm$0.057 & $1.345\pm$0.094 \\*

& & & FWHM & & 191.1$\pm$5 & 245$\pm$45 & & & 219$\pm$22 & 216$\pm$39 \\*

& & & $cz$ & & $19309\pm$17 & $60\pm$24 & & & $-36\pm$25 & $-9\pm$28 \\

\enddata
\tablenotetext{a}{Slope expressed as power-law index, $\alpha$, where $f_\lambda = C\lambda^\alpha$.}
\tablenotetext{b}{Strictly formal measure of the full-width at half-maximum of the Pa$\alpha$ line, deconvolved by the instrumental resolution.}
\tablenotetext{c}{For each galaxy, the following measures are provided for selected lines: equivalent width (EW) is given in nanometers; FWHM as estimated via line flux and peak amplitude (see text), deconvolved by the instrumental resolution; first moment centroid of the line, expressed as recessional velocity ($cz$) for Pa$\alpha$, and velocity offsets from Pa$\alpha$ for all other lines. The first moment measure looses accuracy for weak lines depending on the spectral range summed, and therefore is of limited use here.}
\end{deluxetable}

\begin{deluxetable}{lcclcccccc}
\tabletypesize{\tiny}
\rotate
\tablewidth{0pt}
\tablecaption{Measured Line Properties for Secondary Nuclei\label{tab:mofosec}}
\tablehead{\colhead{Galaxy} & \colhead{Slope\tablenotemark{a}} & \colhead{Pa$\alpha$ FWHM\tablenotemark{b}} & \colhead{Rows\tablenotemark{c}} & \colhead{He I} & \colhead{Pa$\alpha$} & \colhead{H$_2$ S(3)} & \colhead{$[$Fe II$]$} & \colhead{H$_2$ S(1)} & \colhead{Br$\gamma$} \\}
\startdata \\
IRAS 01521$+$5224N & $-4.18$ & 403 & & & & & & & \\*

& & & EW & \nodata & $1.12\pm$0.6 & \nodata & \nodata & \nodata & \nodata \\*

& & & FWHM & & 190$\pm$230 & & & & \\*

& & & $cz$ & & $23990\pm$170 & & & & \\

IRAS 05246$+$0103W & $2.54$ & 54 & & & & & & & \\*

& & & EW & $0.46\pm$0.16 & $7.93\pm$0.31 & \nodata & \nodata & $0.94\pm$0.29 & \nodata \\*

& & & FWHM & 0$\pm$140 & 72$\pm$19 & & & 60$\pm$100 & \\*

& & & $cz$ & $-60\pm$120 & $29103\pm$23 & & & $126\pm$77 & \\

IRAS 08572$+$3915S & $7.02$ & \nodata & & & & & & & \\*

& & & EW & \nodata & $2.92\pm$0.21 & \nodata & \nodata & \nodata & \nodata \\*

& & & FWHM & & 0$\pm$19 & & & & \\*

& & & $cz$ & & $17465\pm$22 & & & & \\

IRAS 09061$-$1248N & $-7.07$ & \nodata & & & & & & & \\*

& & & EW & \nodata & $5.52\pm$0.59 & $0.84\pm$0.29 & \nodata & $2.13\pm$0.39 & \nodata \\*

& & & FWHM & & 252$\pm$76 & 0$\pm$170 & & 253$\pm$95 & \\*

& & & $cz$ & & $22030\pm$30 & $-60\pm$140 & & $81\pm$61 & \\

IRAS 09583$+$4714SW & $-6.81$ & 186 & & & & & & & \\*

& & & EW & \nodata & $1.97\pm$0.16 & $0.94\pm$0.19 & \nodata & \nodata & \nodata \\*

& & & FWHM & & 246$\pm$41 & 320$\pm$180 & & & \\*

& & & $cz$ & & $25696\pm$31 & $138\pm$76 & & & \\

IRAS 10035$+$4852SW & $-10.31$ & 424 & & & & & & & \\*

& & & EW & $0.59\pm$0.23 & $10.33\pm$0.54 & $1.03\pm$0.13 & \nodata & $1.88\pm$0.14 & $1.36\pm$0.17 \\*

& & & FWHM & 50$\pm$160 & 430$\pm$31 & 320$\pm$110 & & 292$\pm$47 & 212$\pm$67 \\*

& & & $cz$ & $-60\pm$130 & $19443\pm$31 & $-61\pm$54 & & $26\pm$42 & $-17\pm$48 \\

IRAS 10190$+$1322W & $-3.94$ & 265 & & & & & & & \\*

& & & EW & \nodata & $4.51\pm$0.14 & $0.41\pm$0.092 & $0.31\pm$0.11 & $0.64\pm$0.13 & $0.72\pm$0.18 \\*

& & & FWHM & & 271$\pm$26 & 40$\pm$110 & 0$\pm$150 & 63$\pm$89 & 150$\pm$120 \\*

& & & $cz$ & & $23012\pm$21 & $-70\pm$64 & $370\pm$170 & $47\pm$56 & $-28\pm$68 \\

IRAS 12112$+$0305NE & $2.53$ & 196 & & & & & & & \\*

& & & EW & \nodata & $21.49\pm$0.52 & $4.62\pm$0.13 & \nodata & $4.05\pm$0.2 & $3.67\pm$0.39 \\*

& & & FWHM & & 271$\pm$19 & 401$\pm$27 & & 366$\pm$49 & 278$\pm$78 \\*

& & & $cz$ & & $21875\pm$20 & $-110\pm$36 & & $-47\pm$30 & $-12\pm$36 \\

IRAS 14348$-$1447NE & $-2.11$ & 283 & & & & & & & \\*

& & & EW & $2.06\pm$0.3 & $15.41\pm$0.35 & $1.98\pm$0.15 & \nodata & $2.5\pm$0.25 & $1.64\pm$0.5 \\*

& & & FWHM & 450$\pm$160 & 272$\pm$14 & 218$\pm$63 & & 274$\pm$76 & 0$\pm$120 \\*

& & & $cz$ & $133\pm$64 & $24681\pm$23 & $-31\pm$35 & & $-66\pm$36 & $20\pm$100 \\

IRAS 15245$+$1019W & $-5.05$ & 426 & & & & & & & \\*

& & & EW & \nodata & $4.36\pm$0.38 & $2.7\pm$0.19 & \nodata & $2.19\pm$0.24 & \nodata \\*

& & & FWHM & & 310$\pm$70 & 406$\pm$74 & & 78$\pm$54 & \\*

& & & $cz$ & & $22577\pm$31 & $-72\pm$42 & & $-100\pm$41 & \\

IRAS 17028$+$5817E & $-1.31$ & 0 & & & & & & & \\*

& & & EW & \nodata & $9.66\pm$0.76 & \nodata & \nodata & \nodata & \nodata \\*

& & & FWHM & & 97$\pm$33 & & & & \\*

& & & $cz$ & & $31938\pm$32 & & & & \\

IRAS 22491$-$1808W & $-8.88$ & 86 & & & & & & & \\*

& & & EW & $0.93\pm$0.38 & $5.59\pm$0.2 & $0.52\pm$0.21 & \nodata & \nodata & \nodata \\*

& & & FWHM & 0$\pm$140 & 0$\pm$23 & 0$\pm$99 & & & \\*

& & & $cz$ & $80\pm$100 & $23274\pm$23 & $-180\pm$160 & & & \\

\enddata
\tablecomments{See notes to Table \ref{tab:mofoprim} for details.}
\end{deluxetable}

Figure~\ref{fig:twod} presents the two-dimensional spectra around the Pa\( \alpha  \)
line for each of the sample galaxies. In each, the horizontal axis origin is
located at the center of the primary nucleus. The apertures corresponding to
the extractions displayed in Figures~\ref{fig:primary} and \ref{fig:secondary}
are indicated at the top and bottom of each panel. Annotations to the left and
right of each panel denote the sense of the slit orientation, the exact position
angles for which are listed in Table~\ref{tab:obs}. The two-dimensional spectrum
for IRAS~09583\( + \)4714 has been split into two panels, though the horizontal
scale reflects the real separation between the two nuclear components.

\subsection{Hydrogen Recombination \& Extinction}

The \( K \) band spectra of the galaxies in this sample are dominated by the
Pa\( \alpha  \) line, which is the strongest line accessible to search for
signs of broad emission. Only in two of the sample galaxies, IRAS~08311\( - \)2459
and IRAS~15462\( - \)0450, is any such emission apparent. Both of these galaxies
are optically classified as Seyfert galaxies, so it seems that obscured AGN
in ULIRGs are either extremely rare, or suffer a few magnitudes of extinction
even at near-infrared wavelengths, as was also suggested by \citet{goldader}.
A few galaxies, such as IRAS~14348\( - \)1447 and IRAS~14394\( + \)5332
have excess emission on the blue side of the Pa\( \alpha  \) line, but in both
cases there appears to be emission from \ion{He}{1} contributing significantly
to this flux, and no appreciable emission on the red side of the narrow Pa\( \alpha  \)
profile. Furthermore, the H\( _{2} \) lines in IRAS~14394\( + \)5332 have
an asymmetric blue shape more typical of wind or outflow phenomena than of velocity-broadening
due to an AGN. IRAS~20087\( - \)0308 has some hint of broad emission at the
base of the Pa\( \alpha  \) line, though not significantly elevated out of
the noise, and again confused by \ion{He}{1}. As seen in Figure~\ref{fig:twod},
the moderate 600 km~s\( ^{-1} \) width of the narrow portion of the Pa\( \alpha  \)
profile in IRAS~20087\( - \)0308 can be attributed to a steep rotation gradient
on the nucleus. The apparent broad base on the Pa\( \alpha  \) line in the
southwestern nucleus of IRAS~12112\( + \)0305 may be spurious, as this is
the one galaxy in the sample whose spectra were not calibrated by a G star observation,
but rather by a template atmospheric spectrum. The excess blue emission happens
to be coincident with the significant CO\( _{2} \) atmospheric absorption at
2.00--2.02 \( \mu  \)m, so that a slight error in estimated column depth results
in a large localized continuum offset in the corrected spectrum.

\begin{figure}
{\par\centering \includegraphics{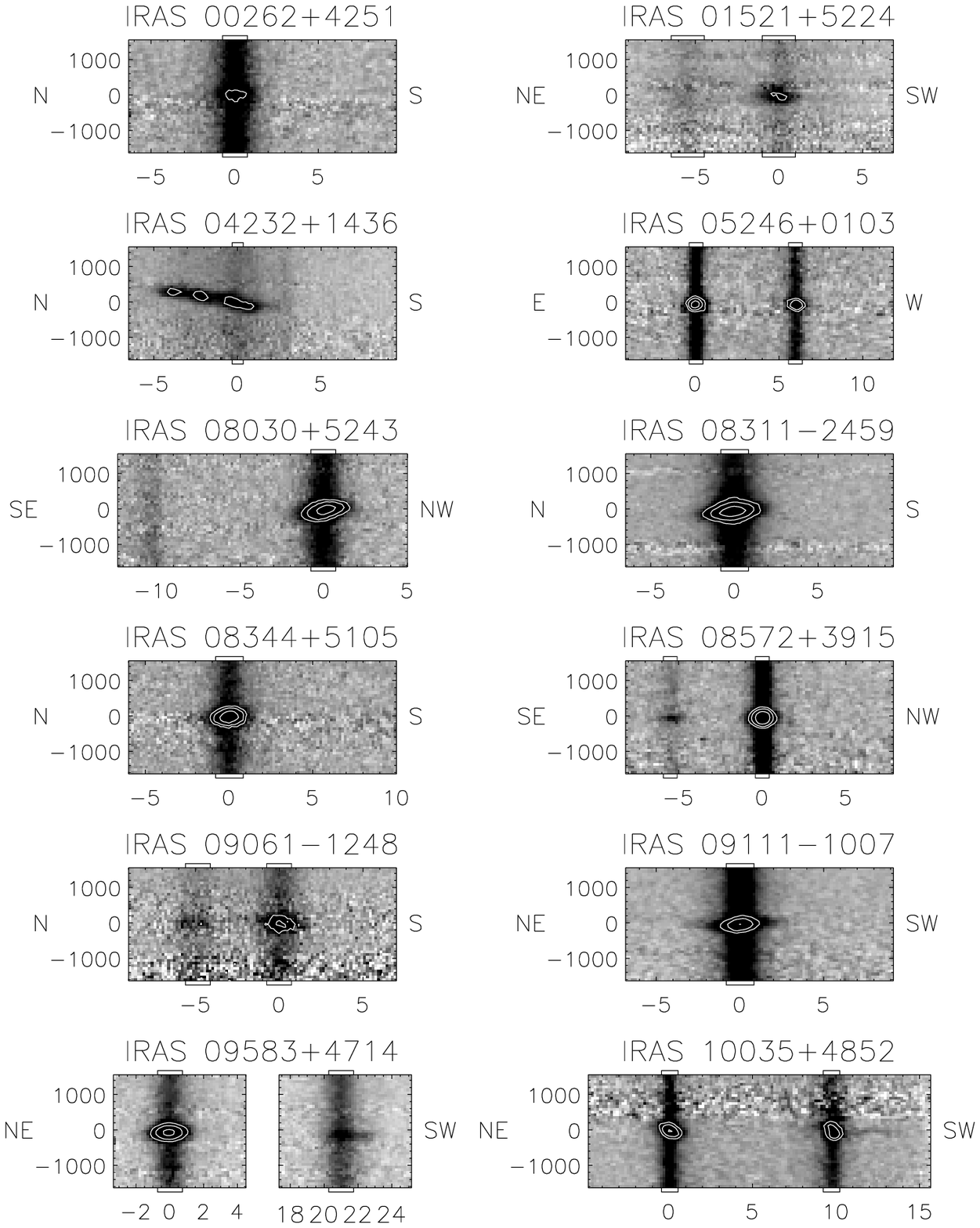} \par}
\caption{\label{fig:twod}Two dimensional spectra around the Pa\protect\( \alpha \protect \)
line for each galaxy. The horizontal scale is in arcseconds, with the primary
nucleus at the origin. The spectral direction spans about 11 resolution elements,
or about 3300 km~s\protect\( ^{-1}\protect \). Extraction apertures are indicated
on the top and bottom of each frame.}
\end{figure}
\setcounter{figcontinue}{\value{figure}}
\begin{figure}
\setcounter{figure}{\value{figcontinue}}
{\par\centering \includegraphics{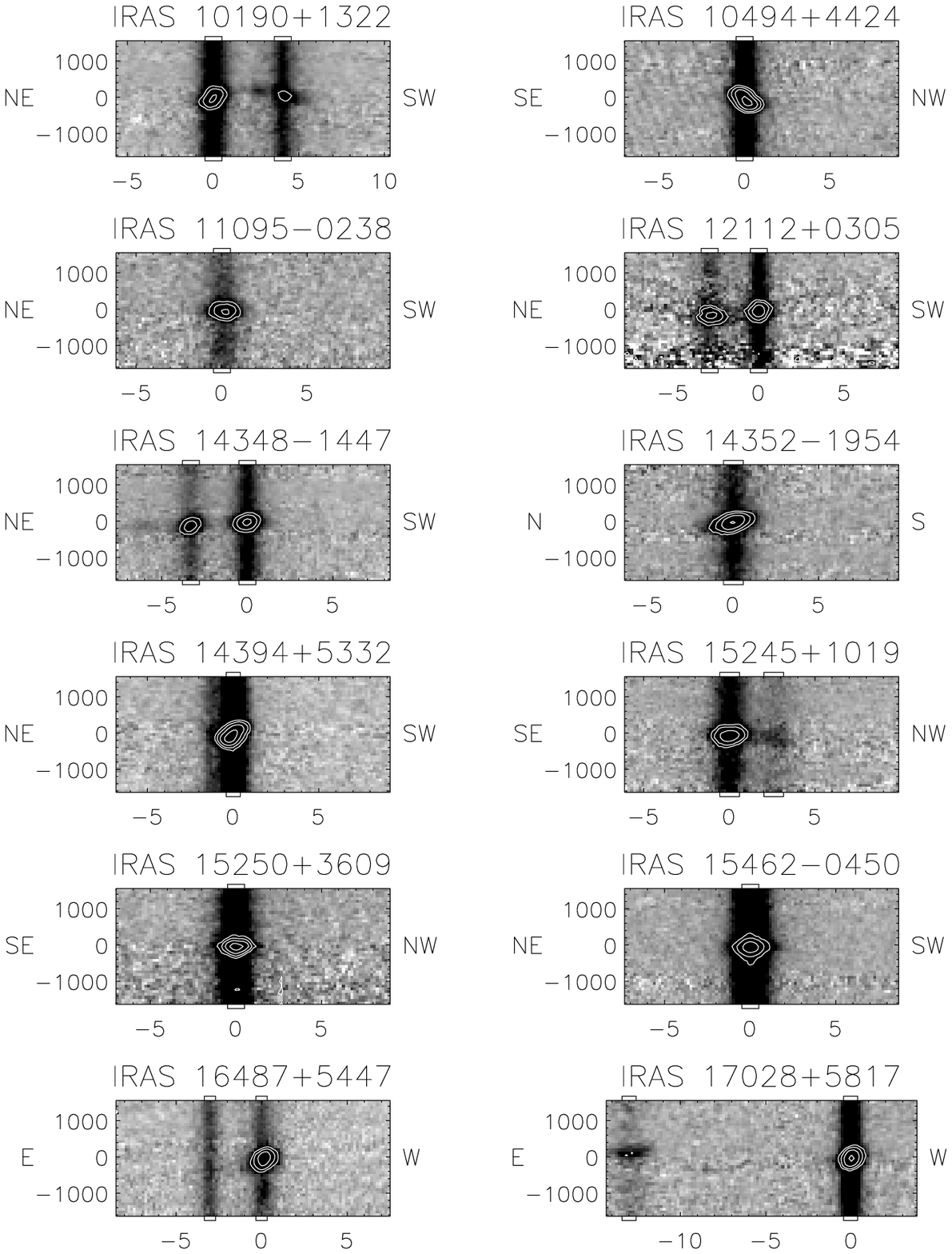} \par}
\mycaption{Figure \the\value{figure}}{continued}
\end{figure}
\begin{figure}
\setcounter{figure}{\value{figcontinue}}
{\par\centering \includegraphics{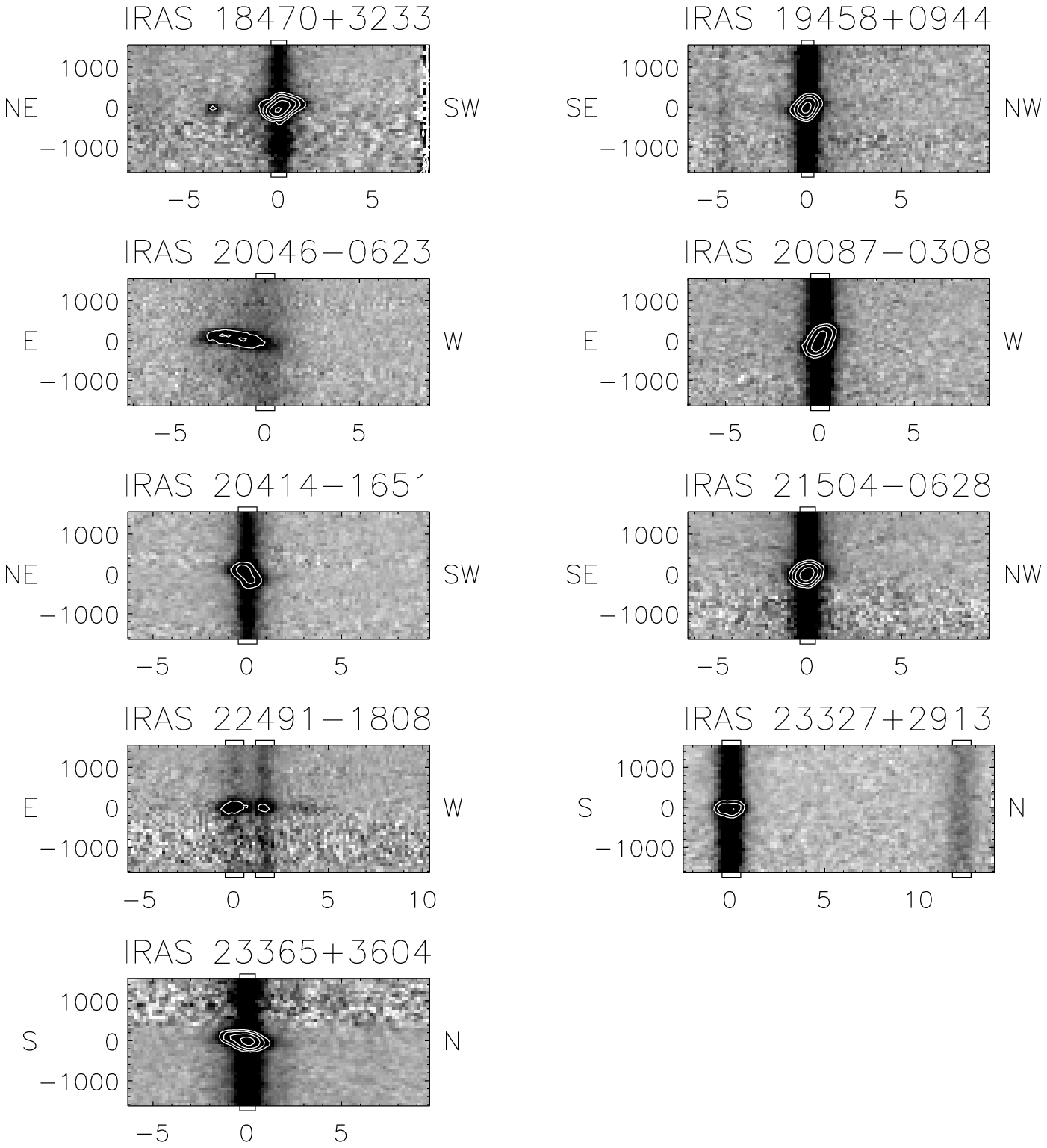} \par}
\mycaption{Figure \the\value{figure}}{continued}
\end{figure}

The two-dimensional spectra of Pa\( \alpha  \) in Figure~\ref{fig:twod} demonstrate
that the atomic recombination line emission in ULIRGs is almost always dominated
by a nuclear component. Several ULIRGs show off-nuclear emission at lower levels,
and a couple of sources appear to be primarily characterized by diffuse line
emission---namely IRAS~04232\( + \)1436 and IRAS~20046\( - \)0623. When
two nuclei are present, the secondary nucleus generally shows emission in Pa\( \alpha  \)
as well. Exceptions to this are IRAS~16487\( + \)5447 and IRAS~23327\( + \)2913.
The extra continuum object seen in the panel with IRAS~08030\( + \)5243 is
a nearby star.

The fraction of the total line emission that is unresolved can be estimated
for each ULIRG by computing the flux contribution from a Gaussian spatial profile
having the width of the estimated seeing and sharing the maximum amplitude of
the spatial line emission distribution. Comparing this flux with the measured
flux provides an upper limit to the fraction of light contained within an unresolved
core. Most of the primary ULIRG nuclei are consistent with being unresolved,
with a median seeing of 1\farcs 0 and a median physical resolution of 1.4~kpc.
The scatter in the estimated fraction of unresolved flux is at the \( \pm  \)20\%
level, mostly due to uncertainty in the seeing estimate. Only 8 of the 33 ULIRGs
have estimated unresolved contributions lower than 80\%, such that the majority
of ULIRGs contain most---if not all---of their Pa\( \alpha  \) emission within
the central 1~kpc radius. Only three ULIRGs in the sample (IRAS~04232\( + \)1436,
IRAS~18470\( + \)3233, and IRAS~20046\( - \)0623) have less than 50\% of
their Pa\( \alpha  \) flux in an unresolved core.

The sample was chosen to allow coverage of both the Pa\( \alpha  \) and Br\( \gamma  \)
emission lines so that a comparison between these lines could be made in order
to assess extinction to the line emitting source. Table~\ref{tab:props} lists
the inferred visual extinction values for the sample galaxies. These estimates
assume an intrinsic ratio for Pa\( \alpha / \)Br\( \gamma  \) of 12.1 \citep[][Case B with $n_e=10^4$ cm$^{-3}$ and $T=10000$ K]{ost},
and \( A_{Pa\alpha }=0.145A_{V} \) and \( A_{Br\gamma }=0.116A_{V} \) \citep[interpolated from the extinction law of][]{extinct}.

Though there exists very little leverage in wavelength between Pa\( \alpha  \)
and Br\( \gamma  \), some ULIRGs are apparently obscured enough to show definite
reddening across this short range. Typical 1\( \sigma  \) errors on the extinction
estimates are around 5 magnitudes, though the actual errors could be larger
if disparate aperture sizes were used in the spectral extractions. Indeed, each
of the four ULIRGs with extinctions more negative than 1\( \sigma  \) have
aperture differences between the Pa\( \alpha  \) and Br\( \gamma  \) spectra
of at least 50\%. Almost half of the ULIRGs have estimated extinctions greater
than the 1\( \sigma  \) limit, and 11 have at least 2\( \sigma  \) extinction
estimates---typically corresponding to \( A_{V}>10 \) mag, with many in the
range of \( A_{V}\sim 25 \) mag.

\begin{figure}
{\par\centering \includegraphics{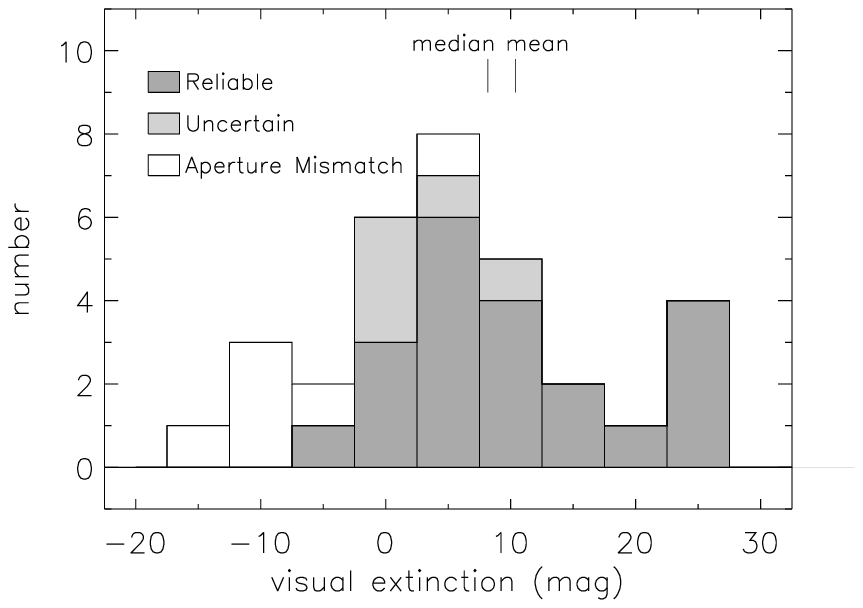} \par}

\caption{\label{fig:extinct}Histogram of estimated visual extinctions to the Pa\protect\( \alpha \protect \)
line emission for the 33 primary ULIRG nuclei. Different shadings represent
varying levels of reliability, with most of the negative estimates owing to
mismatched seeing conditions and therefore significant extraction aperture differences.
The mean and median points are marked for the reliable set of measurements.
One point, corresponding to \protect\( A_{V}\sim 50\protect \), is beyond the
plot boundary.}
\end{figure}

Figure~\ref{fig:extinct} presents the extinction measurements in histogram
form. Different shading represents varying levels of confidence in the measurement.
Higher extinction estimates are allowed greater variance in classifying their
reliability, provided they are significantly different from zero extinction.
Counting only the 22 ULIRGs with reliable extinction measures, the median visual
extinction of \( \sim 8 \) mag corresponds to an extinction at Pa\( \alpha  \)
of approximately one magnitude. This is a lower limit to the nuclear extinction
because we can only estimate extinction to the gas we actually detect at 2 \( \mu  \)m,
which very likely is peripheral to the nuclear region. Nonetheless, one magnitude
of extinction at Pa\( \alpha  \) would not readily hide the presence of a central
AGN. The very large visual extinctions of 25 mag seen in some ULIRGs translate
to almost 4 mag of extinction at Pa\( \alpha  \), which almost certainly would
hide an AGN from view with the currently achieved signal-to-noise ratio in the
ULIRG spectra.

\begin{deluxetable}{lcccc} 
\tabletypesize{\footnotesize}
\tablewidth{0pt}
\tablecaption{Derived Properties\label{tab:props}}
\tablehead{\colhead{Galaxy} & \colhead{Extinction\tablenotemark{a}} & \multicolumn{3}{c}{H$_2$ Line Parameters} \\
& & \colhead{$T_{rot}$} & \colhead{$T_{vib}$\tablenotemark{b}} & \colhead{H$_2$ Excitation} \\
& \colhead{($A_V$)} & \colhead{(K)} & \colhead{(K)} & \colhead{Mechanism\tablenotemark{c}}\\ }
\startdata 

IRAS 00262$+$4251 & \phs\phn\phn\phd$0\pm66$\phd & $1300\pm75$\phn & $4000\pm1000$ & mixture\phm{$\ast$} \\ 

IRAS 01521$+$5224 & \phs\phn\phn\phd$3\pm14$\phd & $1400\pm300$ & 5000 & thermal$\ast$ \\

IRAS 04232$+$1436 & \phs\phn$2.3\pm7.2$ & $2100\pm225$ & & thermal\phm{$\ast$}\\

IRAS 05246$+$0103 & \phs\phn\phn\phd$1\pm15$\phd & $2500\pm375$ & & thermal\phm{$\ast$}\\

IRAS 08030$+$5243 & \phs\phn$0.9\pm4.6$ & $2100\pm450$ & & thermal\phm{$\ast$}\\

IRAS 08311$-$2459 & $-14.5\pm4.3$\tablenotemark{d} &$4300\pm500$ & & thermal\phm{$\ast$}\\

IRAS 08344$+$5105 & \phn$-5.0\pm5.8$ & $1800\pm200$ & $4500\pm2000$ & mixture\phm{$\ast$}\\

IRAS 08572$+$3915 & \phs$12.1\pm3.6$ & $2500\pm300$ & $5000\pm1500$ & mixture\phm{$\ast$}\\

IRAS 09061$-$1248 & \phs\phn$3.9\pm6.1$ & $2200\pm250$ & 3500 & thermal$\ast$ \\

IRAS 09111$-$1007 & \phs\phn$8.7\pm4.0$ & $2300\pm200$ & 2700 & thermal$\ast$ \\

IRAS 09583$+$4714 & \phn$-6.2\pm5.1$\tablenotemark{d} & $2000\pm250$ & & thermal\phm{$\ast$}\\

IRAS 10035$+$4852 & \phn$-9.9\pm5.3$\tablenotemark{d} & $2900\pm300$ & & thermal\phm{$\ast$}\\

IRAS 10190$+$1322 & \phs\phn$4.4\pm3.9$ & $2200\pm125$ & & thermal\phm{$\ast$}\\

IRAS 10494$+$4424 & $-10.8\pm3.4$\tablenotemark{d} &$2400\pm250$ & 3500 & thermal$\ast$ \\

IRAS 11095$-$0238 & \phs\phn$9.5\pm16$\phd & $1700\pm150$ & $3800\pm1300$ & mixture\phm{$\ast$}\\

IRAS 12112$+$0305 & \phs$19.5\pm3.7$ & $2500\pm300$ & 5000 & thermal$\ast$ \\

IRAS 14348$-$1447 & \phs\phn$5.7\pm5.7$ & $2000\pm125$ & & thermal\phm{$\ast$}\\

IRAS 14352$-$1954 & \phs\phn$2.8\pm3.2$ & $1600\pm150$ & & thermal\phm{$\ast$}\\

IRAS 14394$+$5332 & \phs$12.0\pm4.8$ & $2200\pm150$ & 2800 & thermal$\ast$ \\

IRAS 15245$+$1019 & \phs\phn$7.3\pm2.3$ & $4900\pm800$ & & thermal\phm{$\ast$}\\

IRAS 15250$+$3609 & \phs\phn$4.6\pm2.1$\tablenotemark{d} & $2000\pm150$ & $3100\pm600$\phn & mixture\phm{$\ast$}\\

IRAS 15462$-$0450 & \phs\phn$8.2\pm4.4$ & $>4000$ & & thermal\phm{$\ast$}\\

IRAS 16487$+$5447 & \phs$23.3\pm9.9$ & $1700\pm150$ & & thermal\phm{$\ast$} \\

IRAS 17028$+$5817 & \phs$27.2\pm4.4$ & $2200\pm200$ & & thermal\phm{$\ast$}\\

IRAS 18470$+$3233 & \phs\phn$0.5\pm2.7$ & $1600\pm100$ & & thermal\phm{$\ast$}\\

IRAS 19458$+$0944 & \phs$24.3\pm6.3$ & $2300\pm100$ & & thermal\phm{$\ast$} \\

IRAS 20046$-$0623 & \phs\phd\phn$50\pm11$\phd & $1900\pm300$ & & thermal\phm{$\ast$} \\

IRAS 20087$-$0308 & \phd\phn$-10\pm15$\phd & $2200\pm350$ & & thermal\phm{$\ast$} \\

IRAS 20414$-$1651 & \phs\phd\phn$26\pm11$\phd & $2400\pm225$ & & thermal\phm{$\ast$} \\

IRAS 21504$-$0628 & \phs\phn$3.4\pm2.8$ & $2600\pm300$ & & thermal\phm{$\ast$} \\

IRAS 22491$-$1808 & \phd\phn$16\pm11$\phd & $2100\pm250$ & & thermal\phm{$\ast$} \\

IRAS 23327$+$2913 & \phs\phn\phn\phd$2\pm73$\phd & $1600\pm200$ & & thermal\phm{$\ast$} \\

IRAS 23365$+$3604 & \phs$14.5\pm2.6$ & $1900\pm150$ & $4200\pm900$ & mixture\phm{$\ast$} \\

\enddata
\tablenotetext{a}{Based on Pa$\alpha /$Br$\gamma$ line ratio.}
\tablenotetext{b}{Given only for definite thermal/fluorescent mixtures (with errors) and for likely mixtures (when both of the observed 2--1 lines are above the rotation temperature prediction). The vibration temperature assumes an ortho-to-para ratio of 2:1 in the 2--1 transition states.} 
\tablenotetext{c}{Classified as consistent with purely thermal, definite mixture of thermal and fluorescence, or possible mixture of the two (with asterisk).} 
\tablenotetext{d}{Apertures for the Pa$\alpha$ and Br$\gamma$ spectral extractions differ by at least 50\%.}
\end{deluxetable}

Because an active galactic nucleus is believed to indicate the presence of a
QSO (quasi-stellar object) in the galaxy nucleus, one may quantify the amount
of obscuration necessary to hide a central AGN by estimating the extinction
required to hide the line flux expected from the high-velocity wings of Pa\( \alpha  \)
in a typical QSO of the appropriate bolometric luminosity. For this purpose,
we use the median QSO spectrum presented in \citet{twm99} to represent the appearance
of an unobscured quasar. Identifying the region around 1.881 \( \mu  \)m (\( \approx 1000 \)
km~s\( ^{-1} \) redward of the Pa\( \alpha  \) line center) as being well
away from the narrow Pa\( \alpha  \) emission, we measure the median QSO flux
density to be 22\% higher than the QSO continuum level. It is found that the
broad Pa\( \alpha  \) emission from an exposed QSO contributing half the bolometric
luminosity of a ULIRG would be comparable at 1.881 \( \mu  \)m to the typical
ULIRG continuum emission (neglecting the contribution from the QSO continuum).
Therefore, hiding the presence of the broad emission requires an attenuation
factor that is comparable to the signal-to-noise ratio of the ULIRG continuum.
In the case of the median ULIRG spectrum presented in \citeauthor{twm99}, this
translates to an extinction at Pa\( \alpha  \) of 4.4 mag, and therefore a
visual extinction, \( A_{V}\approx 30 \) mag. The spectra of the individual
ULIRGs presented here typically have continuum signal-to-noise ratios around
20, so that \( A_{Pa\alpha }=3.1 \) mag, or \( A_{V}=20 \)--25 mag of extinction
is required to hide a QSO at half the ULIRG luminosity. Details on this calculation
appear in Appendix~\ref{qsoextinct}.

The estimates of extinction made in this way are conservative in two regards.
First, the aperture photometry of ULIRGs used to establish typical continuum
flux densities are based on 5\( '' \) apertures, such that extra flux is included
from diffuse emission away from the nucleus. In smaller apertures corresponding
to the spectral extractions employed in this paper, the ULIRG continuum is diminished,
while the spatially unresolved QSO component would be hardly affected, thus
increasing the QSO-to-ULIRG contrast considerably, making the QSO more difficult
to conceal. The second conservative aspect is that the criterion of forcing
the broad Pa\( \alpha  \) from the QSO to lie at the 1\( \sigma  \) level
on the ULIRG continuum would not hide the broad emission very effectively, the
reason being that the 1\( \sigma  \) level is computed for a single pixel (one-quarter
of a resolution element). When summed over a few resolution elements (i.e.,
up to the narrow shoulder of the Pa\( \alpha  \) line), the broad signal would
stand out at more than the 3\( \sigma  \) level.

The result that, given the current level of sensitivity, at least 25 magnitudes
of visual extinction are required to obscure a QSO in a typical ULIRG, together
with the paucity of detections of low-level broad line emission---especially
evident in the median spectrum of \citet{twm99}---leads us to conclude that
AGN are either rare in ULIRGs, or very deeply buried. There are five ULIRGs
in the present sample with measured extinctions based on the Pa\( \alpha  \)/Br\( \gamma  \)
ratio that are high enough to hide AGN without requiring additional attenuation.
We are unable to determine the likelihood that the rest of the ULIRGs have the
requisite nuclear extinctions to hide the presence of AGN.

The technique of comparing ULIRG spectra to a typical QSO spectrum can also
be used to estimate the extinction to the broad-line region seen in IRAS~15462\( - \)0405.
Based on the flux density at 1.881 \( \mu  \)m, the Pa\( \alpha  \) emission
contributes an additional 9\% to the continuum level, some of which is likely
continuum emission from the QSO itself. Correcting for this, one finds an excess
over the assumed ULIRG continuum (60\% of the total continuum) measuring 15\%,
suggesting an extinction \( A_{Pa\alpha }=2.0\pm 0.8 \) mag, or \( A_{V}=14\pm 6 \)
mag.

\subsection{Molecular Hydrogen Emission}

\begin{figure}
{\par\centering \includegraphics{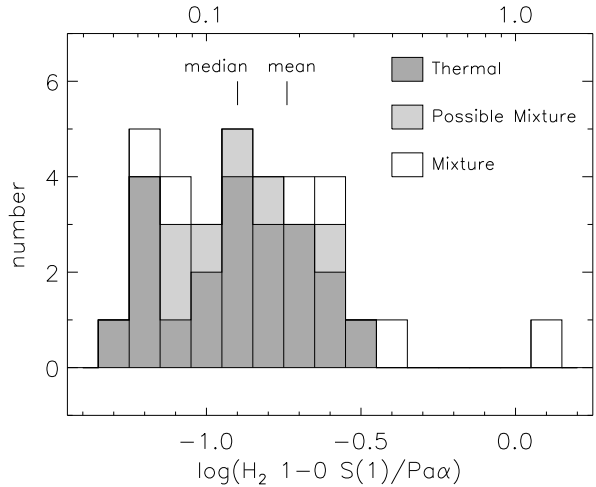} \par}

\caption{\label{fig:pah2}Ratio of the H\protect\( _{2}\protect \)~1--0 S(1) line
strength to that of Pa\protect\( \alpha \protect \) for the 33 primary ULIRG
nuclei. A linear scale appears at top. Typical 1\protect\( \sigma \protect \)
errors are comparable to the bin width, though perhaps a little smaller. The
mean is performed on the linear scale, rather than on the logarithmic scale.
Shadings indicating excitation mechanisms are included for reference.}
\end{figure}

The most abundant species of line emission in the ULIRG spectra arise from the
vibration-rotation spectrum of molecular hydrogen. The molecular hydrogen emission
shows up strongly in \emph{all} of the primary nuclei of this sample. Three
or four lines are typically present in each spectrum, most notably the odd rotation
state 1--0 transitions. The H\( _{2} \) line strengths vary by more than one
order-of-magnitude relative to Pa\( \alpha  \). Figure~\ref{fig:pah2} shows
the range of observed line ratios, comparing the H\( _{2} \)~1--0 S(1) line
to Pa\( \alpha  \). A typical ULIRG has a H\( _{2} \)~1--0 S(1)\( / \)Pa\( \alpha  \)
ratio around 0.15, varying from \( \sim 0.05 \) to 0.40. One galaxy, IRAS~00262\( + \)4251
exhibits H\( _{2} \) emission even stronger than the Pa\( \alpha  \) line.
The bins in Figure~\ref{fig:pah2} are shaded to reflect the classification
of the H\( _{2} \) excitation mechanisms, as will be discussed below.

The apparent weakness of the even rotation transitions is an indication that
the emission is thermally excited, as the thermal condition produces three times
as many odd (ortho) states as even (para) states. This so-called ortho-to-para
ratio is more typically in the range of 1.0--1.8 for fluorescent excitation
conditions \citep{bvd}. In addition, the lack of strong emission from the 2--1
vibration transition indicates a thermal origin.

\begin{figure}
{\par\centering \includegraphics{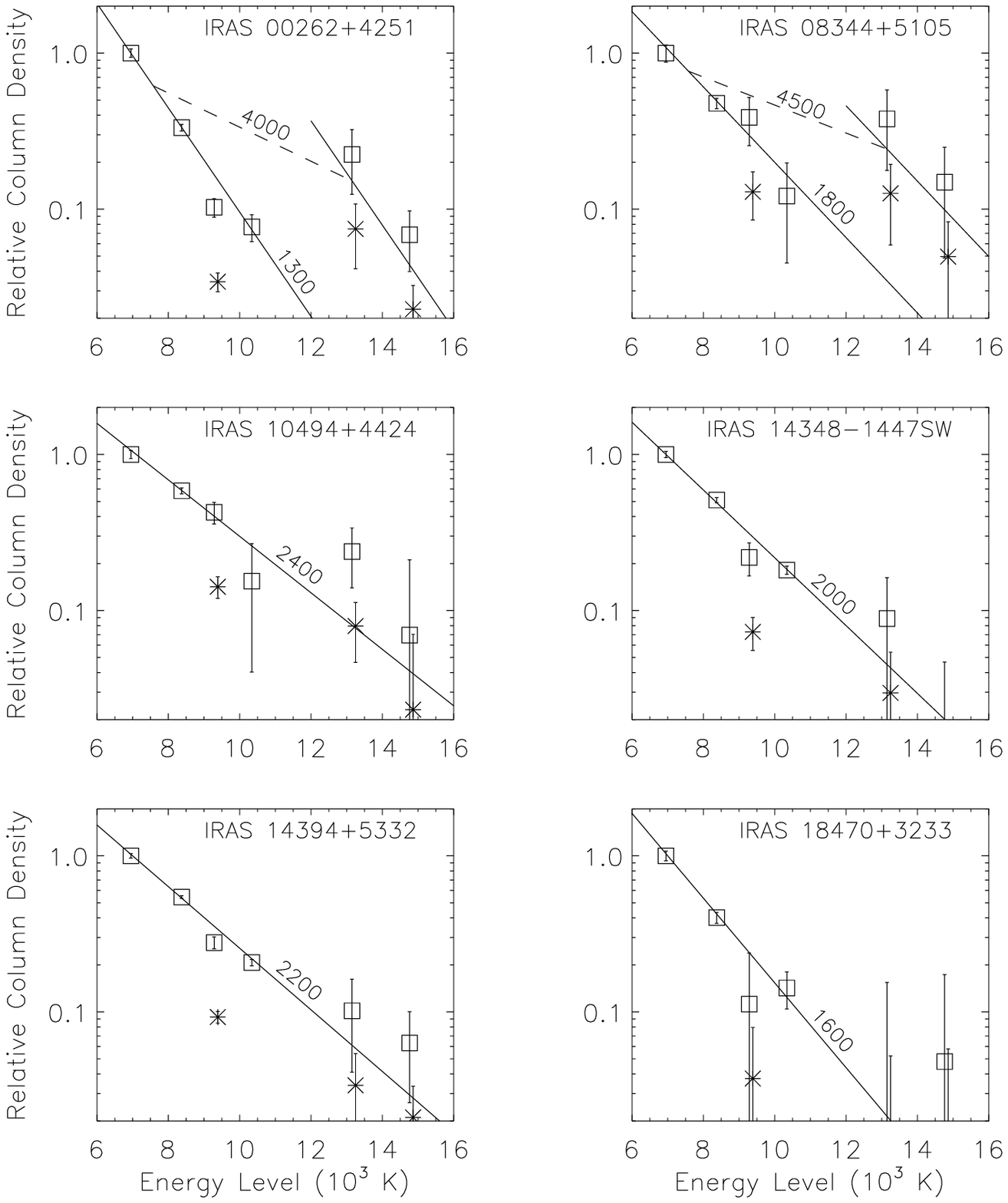} \par}

\caption{\label{fig:h2plots}Molecular hydrogen population diagrams for six of the ULIRGs,
showing the range of properties seen in the H\protect\( _{2}\protect \) excitation.
The top two show examples of mixed thermal and fluorescent excitation, while
the bottom four are consistent with a single temperature characterization. IRAS
10494\protect\( +\protect \)4244 and IRAS 14394\protect\( +\protect \)5332
may have some slight component of fluorescent excitation, judging by the elevated
2--1 lines. For even (para) rotation states, two points are plotted, corresponding
to ortho-to-para ratios of 3:1 and 1:1, with boxes corresponding to the 3:1
case, and asterisks for 1:1. The population diagram for IRAS~18470\protect\( +\protect \)3233
is fairly typical of most of the ULIRGs in this sample in terms of excitation
information and signal-to-noise ratio. Line fits are accompanied by temperatures
corresponding to the slopes.}
\end{figure}

Population diagrams were constructed for each of the primary nucleus extractions
in the sample, and each characterized in terms of rotation temperature, vibration
temperature (where appropriate) and classified in terms of thermal or mixed
thermal/fluorescent excitation. The construction of population diagrams is discussed
in Appendix~\ref{append}. Examples of these diagrams are shown in Figure~\ref{fig:h2plots}.
The even rotation state transitions are plotted as two points representing ortho-to-para
ratios of 3:1 and 1:1. The 3:1 points, represented by boxes in Figure~\ref{fig:h2plots},
are almost always preferred over the 1:1 points (asterisks) appearing below
them. The great majority of ULIRGs in this sample have no detections in the
2--1 transition lines, but with error bars consistent with purely thermal excitation.
Solid lines in Figure~\ref{fig:h2plots} represent rotation temperature fits,
and dashed lines, where present, indicate the vibration temperature separating
1--0 and 2--1 points. A few ULIRGs appear to be definite mixtures of thermal
and fluorescent excitation, with the population diagrams showing the 2--1 data
points well above the rotation temperature line passing through the 1--0 points.
A number of other ULIRGs have somewhat elevated 2--1 measures, with both the
S(2) and S(4) lines above the 1--0 rotation temperature line, but with error
bars that are still consistent with pure thermal emission. The excitation characterizations
are summarized in Table~\ref{tab:props}. For the mixed cases, a vibration temperature
estimate is given, assuming that the high energy states exist with a lower ortho-to-para
ratio, likely around 2:1. This has been observed in other mixed excitation cases
\citep[e.g.,][]{tanaka}, and is explained by the fact that the enhanced 2--1
transition emission arises from molecules that are primarily excited by fluorescent
processes, and thus reflect an ortho-to-para ratio closer to that expected for
pure fluorescence. 

One point that should be mentioned in the context of the line population diagrams
is the possible blend of a 2.0024 \( \mu  \)m {[}\ion{Fe}{2}{]} line with the
H\( _{2} \)~2--1 S(4) line at 2.0035 \( \mu  \)m. The expected flux ratio
between the {[}\ion{Fe}{2}{]} line at 1.9670 \( \mu  \)m and that at 2.0024
\( \mu  \)m is 2--3:1. With the 1.9670 \( \mu  \)m line appearing to be present
in a number of ULIRGs, it is to be expected that the H\( _{2} \)~2--1 S(4)
line will suffer some contamination by the {[}\ion{Fe}{2}{]} line. Indeed, the
population diagrams often show an anomalously high 2--1 S(4) line flux, with
an incompatibly low 2--1 S(2) flux. There is a good correlation between such
cases and galaxies for which the 1.9670 \( \mu  \)m {[}\ion{Fe}{2}{]} line
is believed to be present.

\begin{figure}
{\par\centering \includegraphics{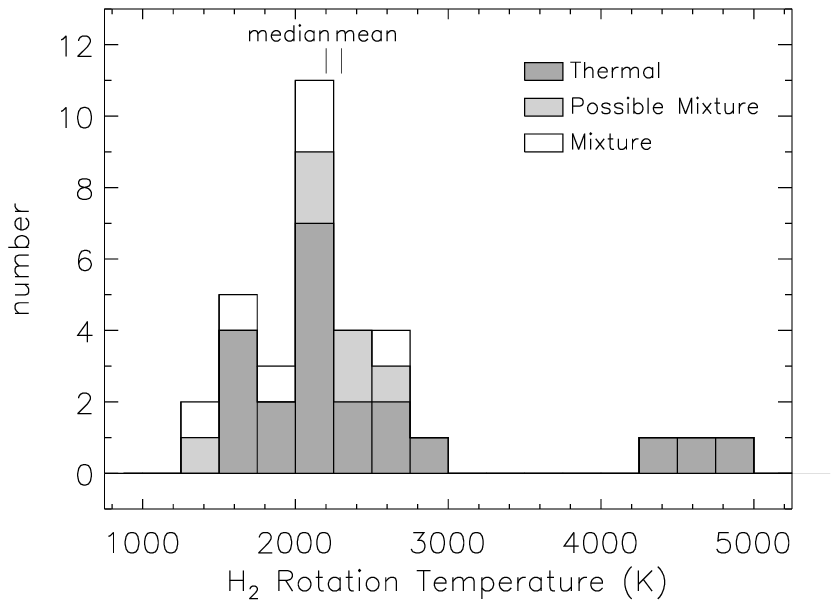} \par}

\caption{\label{fig:h2temp}Rotation temperatures determined for the 33 primary ULIRG
nuclei. The majority are consistent with thermal emission, for which the rotation
temperature represents the ambient equilibrium temperature. A few are definitely
mixtures of thermal and fluorescent processes, with vibration temperatures in
the range of 3100--5000 K.}
\end{figure}

Figure~\ref{fig:h2temp} illustrates the distribution of H\( _{2} \) rotation
temperatures, indicating a strong clustering around 2200 K. Two of the three
ULIRGs with estimated rotation temperatures in excess of 4000 K are classified
as having AGN. It is possible in these cases that a significant fraction of
the H\( _{2} \) emission arises from molecular gas excited by X-ray radiation
from the central source. \citet{draine} found this mechanism to be the most
viable source of H\( _{2} \) excitation in NGC~6240, with the X-rays in this
instance most probably arising from frequent supernovae or from merger-induced
shock activity. Table~\ref{tab:h2} of Appendix~\ref{append} provides expected
H\( _{2} \) line ratios in the thermal regime across the 1.8--2.4 \( \mu  \)m
wavelength range.

Referring to Figure~\ref{fig:pah2}, there does not appear to be an overwhelming
trend between H\( _{2} \) line strength---relative to Pa\( \alpha  \)---and
excitation mechanism, except for a possible slight tendency for mixtures to
be seen among the brighter lines. This could simply be a consequence of higher
signal-to-noise levels for the weak 2--1 lines in those galaxies with bright
H\( _{2} \) lines. More mixed states may be revealed with greater sensitivity,
most likely in those cases identified here as possible mixtures.

\subsection{Other Emission Line Strengths}

The \ion{He}{1} line\footnotemark[1] is detected in the majority (27/33) of
the primary nuclear spectra, and in a number of the secondary spectra. The noise-weighted
mean of detected \ion{He}{1}\( / \)Pa\( \alpha  \) is 0.058, with a relatively
small dispersion of \( \sim 0.03 \) about this value. The median \ion{He}{1}\( / \)Pa\( \alpha  \)
ratio is 0.063 among detected lines, and 0.056 including non-detections as zero-valued
points. The full range of detected ratios goes from 0.02 (IRAS~15245\( + \)3609)
to about 0.11 (IRAS~14348\( - \)1447; IRAS~14394\( + \)5332). Lower resolution
near-infrared spectra could easily mistake this line for a blue wing on the
Pa\( \alpha  \) line. \footnotetext[1]{The \ion{He}{1} emission line just blueward of Pa$\alpha$ is a combination of two lines at 1.86860 and 1.86969
$\mu$m, both of the same $1s4f$--$1s3d$ transition, but the former a triplet state and the latter a singlet. As such, the shorter wavelength line is three times stronger than the longer wavelength line, with a centroid wavelength of 1.8689 $\mu$m. The splitting between lines is less than half a resolution element for the present spectra, so that the line combination is treated as one.}

The {[}\ion{Fe}{2}{]} line at 1.9670 \( \mu  \)m appears in several spectra,
most notably the primary nuclei of IRAS~00262\( + \)4251, IRAS~09111\( - \)1007,
IRAS~15245\( + \)1019, and IRAS~15250\( + \)3609, and the secondary nucleus
of IRAS~10190\( + \)1322. These stronger {[}\ion{Fe}{2}{]} emitters have {[}\ion{Fe}{2}{]}\( / \)Pa\( \alpha  \)
ratios up to 0.05, though IRAS~00262\( + \)4251, with its weak Pa\( \alpha  \),
has a ratio around 0.12. There are many possible line detections near the noise
limit that have typical {[}\ion{Fe}{2}{]}\( / \)Pa\( \alpha  \) ratios around
0.01--0.02. Lines weaker than 1\% of Pa\( \alpha  \) are generally too weak
to be detected with confidence. The {[}\ion{Fe}{2}{]} line reported for IRAS~08311\( - \)2459
is possibly confused with the wing of {[}\ion{Si}{6}{]}. A more thorough analysis
of this line complex at slightly higher resolution does not find conclusive
evidence for the {[}\ion{Fe}{2}{]} line in this galaxy \citep{twm0831}.

Searching for the 1.9629 \( \mu  \)m {[}\ion{Si}{6}{]} line was a primary motivation
for the present spectroscopic survey. With a 167 eV excitation potential, this
line indicates a high probability of the presence of an active nucleus. While
it is possible to produce Si\( ^{5+} \) atoms via powerful shocks, \citet{marconi}
find that photoionization is the more likely excitation mechanism among a sample
of nearby Seyfert galaxies. As stars are incapable of producing significant
quantities of photons at this high energy, the natural photoionization source
is an active nucleus with its power-law spectrum extending into the far-ultraviolet
wavelength range.

Only one galaxy in the sample of 33 has prominent {[}\ion{Si}{6}{]} emission.
IRAS~08311\( - \)2459 has a very strong {[}\ion{Si}{6}{]} line, comparable
in strength to the adjacent H\( _{2} \)~1--0 S(3) line, and appearing broader
than the other lines in the spectrum. A detailed discussion of the {[}\ion{Si}{6}{]}
line in this galaxy can be found in \citet{twm0831} and \citet{twm99}. 

The only other galaxy in the sample with possible {[}\ion{Si}{6}{]} emission
is IRAS~05246\( + \)0103. The spectrum of this galaxy shows emission at the
expected position of {[}\ion{Si}{6}{]}, with a signal-to-noise ratio of \( \sim  \)5.
The adjacent H\( _{2} \) line appears broad on the short-wavelength side as
well, calling into question the nature of the emission on the long-wavelength
side of this line. No emission appears at this position in the secondary nucleus
spectrum of Figure~\ref{fig:secondary}, ruling out a problem with atmospheric
calibration or related issues. The H\( _{2} \) 1--0 S(1) line appears unresolved
in the primary nucleus, strongly suggesting that the {[}\ion{Si}{6}{]} emission
is real. While IRAS~05246\( + \)0103 has no evidence of broad line emission,
it does have a peculiar radio spectrum. \citet{crawford} find this galaxy to
be a gigahertz-peaked-spectrum (GPS) radio source, with a 6 cm flux density
much higher than that measured at 20 cm. This condition is not found in many
ULIRGs, with the Parks (PKS) galaxy IRAS~13451\( + \)1232 being the only other
2 Jy ULIRG identified as a GPS source. This galaxy may also have {[}\ion{Si}{6}{]}
emission, as reported by \citet{vex97}, though the blend with the H\( _{2} \)
line at their lower resolution makes it difficult to estimate the relative contribution
of {[}\ion{Si}{6}{]}. The rest of the H\( _{2} \) lines in IRAS~13451\( + \)1232,
as presented in \citeauthor{vex97}, indicate a H\( _{2} \) rotation temperature
around 2500 K, in which case the {[}\ion{Si}{6}{]} line would account for \( \sim  \)30\%
of the total combined line flux. IRAS~13451\( + \)1232 does show impressive
broadening of the Pa\( \alpha  \) line, clearly standing out as an AGN-dominated
source. While IRAS~05246\( + \)0103 has a similar radio spectrum to IRAS~13451\( + \)1232,
it is almost three orders of magnitude less luminous at 20 cm, according to
\citeauthor{crawford} Perhaps there is a low-luminosity active nucleus component
to IRAS~05246\( + \)0103, with the total infrared luminosity dominated by
star formation.

\subsection{Relative Nuclear Velocities}

The double nucleus ULIRGs often exhibit strong Pa\( \alpha  \) emission from
both nuclei, which can be used to study orbital dynamics of the galactic pair.
Table~\ref{tab:vels} lists the relative radial velocities for the 12 ULIRGs
with measured Pa\( \alpha  \) velocities on both nuclei. The reported errors
are considerably smaller than would be obtained by simply subtracting the velocities
presented in Table~\ref{tab:mofosec} from those in Table~\ref{tab:mofoprim},
reflecting the fact that the relative velocities are not affected by systematic
errors associated with wavelength calibration or slit illumination. 

\begin{deluxetable}{lcc} 
\tabletypesize{\footnotesize}
\tablewidth{0pt}
\tablecaption{Relative Nuclear Radial Velocities\label{tab:vels}}
\tablehead{\colhead{Galaxy} & \colhead{Relative Velocity}& \colhead{Projected Separation} \\
&\colhead{(km s$^{-1}$)} & \colhead{(kpc)}\\}
\startdata 

IRAS 01521$+$5224 & \phn$-50\pm25$\tablenotemark{a} & \phn7.6 \\

IRAS 05246$+$0103 & \phn\phn$-1\pm12$ & \phn9.8\\

IRAS 08572$+$3915 & \phn$-26\pm11$ & \phn5.8 \\

IRAS 09061$-$1248 & \phs\phn$12\pm28$ & \phn7.2 \\

IRAS 09583$+$4714 & \phn$-53\pm30$ & 37.5 \\

IRAS 10035$+$4852 & \phs\phn\phn$8\pm28$ & 11.2 \\

IRAS 10190$+$1322 & \phs$114\pm11$ & \phn5.3 \\

IRAS 12112$+$0305 & \phn$-98\pm13$ & \phn3.7 \\

IRAS 14348$-$1447 & $-112\pm10$ & \phn4.7 \\

IRAS 15245$+$1019 & \phn$-54\pm27$ & \phn3.4 \\

IRAS 17028$+$5817 & \phs$120\pm22$ & 23.0 \\

IRAS 22491$-$1808 & \phn\phn$-3\pm10$ & \phn2.2 \\

\enddata
\tablenotetext{a}{From \citet{twmpifs}. The current dataset yields $-22\pm 169$.}
\end{deluxetable}

\begin{figure}
{\par\centering \includegraphics{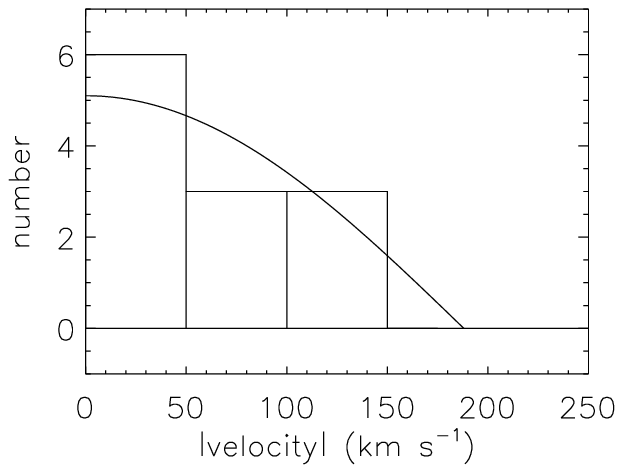} \par}

\caption{\label{fig:velhist}Histogram of the absolute values of projected relative
nuclear velocities. If all ULIRGs had the same relative velocity, projection
effects would result in our observing a cosine distribution. The best fit cosine
to the data has a 190 km~s\protect\( ^{-1}\protect \) velocity amplitude.}
\end{figure}

Surprisingly, the velocity differences are very small, with the largest difference
at 120 km~s\( ^{-1} \). Random projection effects diminish the observed relative
velocities statistically by a factor of two, as the most probable observing
configuration involves pure angular motion on the plane of the sky. If all ULIRGs
have the same relative nuclear velocity, random orientations would produce a
distribution of observed velocities described by a cosine curve, with the peak
at zero observed relative velocity. Binning the data from Table~\ref{tab:vels}
into 50 km~s\( ^{-1} \) bins, the best fit to a cosine function has a maximum
velocity of about 190 km~s\( ^{-1} \), shown in Figure~\ref{fig:velhist}.
Using 40 km~s\( ^{-1} \) bins results in a 165 km~s\( ^{-1} \) maximum relative
velocity.

The lack of high relative velocities among ULIRGs contrasts with the predictions
obtained either from simple Newtonian calculations or from more sophisticated
\( N \)-body models of galaxy interactions which include gas dynamics and dynamical
friction. Below we discuss these two approaches, with attention to how these
models might be modified to match the observations.

First, we will treat the galaxies as point masses approaching each other from
far away. These calculations assume conservative dynamics---i.e., that all loss
of potential energy is converted into kinetic energy of the galaxies. In reality
some fraction of the energy is lost to dynamical friction, which is why the
galaxies do eventually merge. The foregoing discussion will not explicitly treat
the frictional component, but will concentrate on the more basic elements of
merger dynamics that offer much greater latitude in the ultimate relative velocities
calculated. In the point mass scenario, the galaxies would reach a relative
velocity \( \Delta v=2\sqrt{GM/r} \) in excess of 1300 km~s\( ^{-1} \) at
a separation distance of 10 kpc, assuming total system masses of \( 10^{12}M_{\odot } \)
per galaxy. Even ten times less mass per galaxy results in over 400 km~s\( ^{-1} \)
of relative motion at 10 kpc. Changing the initial separation from infinity
to 100 kpc or 50 kpc reduces these speeds by 5\% and 10\%, respectively. 

The high velocities computed above are obviously not consistent with the observations,
but perhaps understandably given that galaxies are not well represented by point
masses at close range. Galaxy masses are dominated by very large dark halos
that act to soften the potential gradients---i.e., accelerations---experienced
during a close encounter.

Dark halos are thought to exist based on the anomalously flat rotation curves
of galaxies to very large radii. Such phenomena suggest the enclosed spherical
mass is proportional to radius, such that the circular rotation velocity, \( v_{c}=\sqrt{GM_{encl}/r} \)
is constant. Rotation curves of large samples of galaxies via optical emission
lines \citep[e.g.,][]{dad} indicate that this scaling persists at least out
to 10--15~kpc. Observations in \ion{H}{1} find a continuation of the flat rotation
curves at radii 2--3 times the optical radii, often to 20--30 kpc \citep[e.g.,][]{casertano,fich}.
Assuming a spherical halo distribution, the total halo mass following the \( M_{encl}\propto r \)
scaling is given by\[
M_{\rm halo}=10^{11}\left( \frac{v_{c}}{208\, \rm km\, s^{-1}}\right) ^{2}\left( \frac{r_{\rm cutoff}}{10\, \rm kpc}\right) M_{\odot },\]
 where \( v_{c} \) is the circular velocity, and \( r_{\rm cutoff} \) is the
outer dimension of the halo distribution. Under this prescription, two galaxies
falling from infinity are moving with a relative velocity of \( \sqrt{2}v_{c} \)
by the time the halos touch, or when the galaxies are \( 2r_{\rm cutoff} \)
apart. The galaxies will clearly accelerate even more as the separation decreases
to \( \sim 10 \) kpc, even though the acceleration is somewhat moderated by
the effect of overlapping halo masses. This already presents a problem in that
the observed relative velocities are less than the typical rotation velocities
observed in ULIRGs.

\begin{figure}
{\par\centering \includegraphics{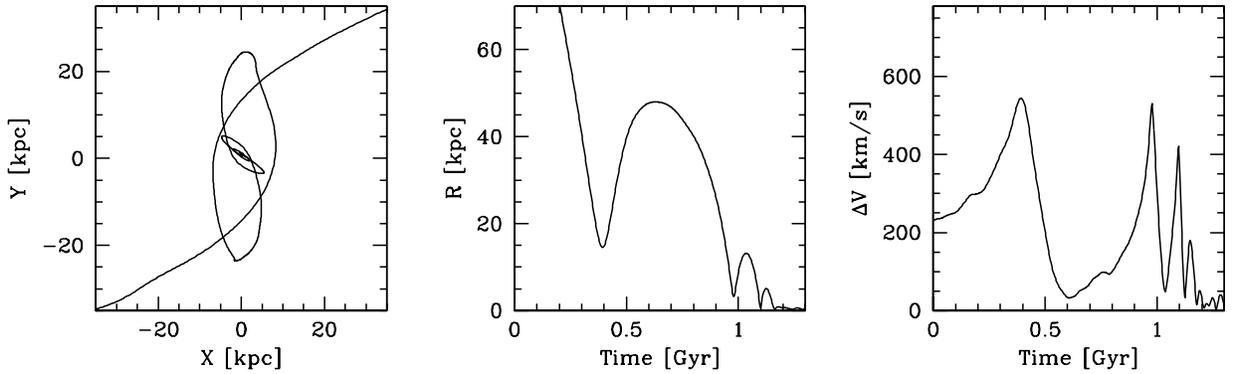} \par}

\caption{\label{fig:mihos}Simulations of galaxy interactions, courtesy of J. C. Mihos.
At left is the orbital track, at center is nuclear separation as a function
of time, and at right is the relative velocity as a function of time. The model
galaxies each have a total mass of \protect\( 4\times 10^{11}\, M_{\odot }\protect \),
and possess Milky-Way-like disk rotation velocities. }
\end{figure}

The simulations by \citet{hos96} provide a means of evaluating encounter velocities
via models employing realistic halo potentials and dynamical friction. Output
from these models is shown in Figure~\ref{fig:mihos}. The model defines a
halo that yields a total galaxy mass of \( 4\times 10^{11} \) \( M_{\odot } \)
when scaled to the Milky Way's rotation velocity. The halo has a relatively
small physical extent, containing most of its mass within 25 kpc, and with an
effective cutoff radius at 40 kpc. The simulations start with identical galaxies
approaching from infinity (parabolic orbits), resulting in time-averaged velocities
of \( \sim 350 \) km~s\( ^{-1} \) when the nuclei are between 4--25 kpc apart.
This average sums over first, second, and third encounters. The peak velocities
are in the neighborhood of 500 km~s\( ^{-1} \), as seen in Figure~\ref{fig:mihos}.
Starting the galaxies from rest at a 100 kpc separation rather than infinity
reduces the peak velocities by only \( \sim 15 \)\%. 

Many ULIRGs are found at times well after the first major encounter, judging
by the fact that their tidal features, formed at the epoch of the first significant
encounter, are often vastly more extended than the nuclei are separated. However,
some ULIRGs are believed to be experiencing their first major encounter \citep{twmpifs}.
Looking at the first-encounter and late-encounter model predictions separately,
the simulation discussed above predicts first-encounter velocities averaging
475 km~s\( ^{-1} \), after which the galaxies separate by 50 kpc, followed
by a final merger sequence with an average velocity around 275 km~s\( ^{-1} \),
ranging from 0--500 km~s\( ^{-1} \) during this time.

Models with larger-scale halos maintaining the same internal rotation structure
demand higher total masses, resulting in increased net velocity differences
between the merging galaxies. On the other hand, if ULIRGs lacked any halo material
past 10 kpc, corresponding to total masses around \( 10^{11} \) \( M_{_{\odot }} \),
the galaxies would behave much more like point masses, and dynamical friction
would be less important. In this scenario, the 10 kpc velocities would be approximately
400 km~s\( ^{-1} \).

It is very difficult to reproduce the small observed relative velocities among
ULIRGs with simple physical models---with or without large halos. ULIRGs are
massive, gas rich galaxies with rotation velocities and near-infrared absolute
magnitudes characteristic of galaxies at least as massive as the Milky Way.
The highest observed relative velocity in this sample is a mere 120 km~s\( ^{-1} \),
though the observed distribution is consistent with a typical deprojected velocity
difference of 200 km~s\( ^{-1} \). Some ULIRGs have been observed with velocity
differences very near this value \citep[e.g., IRAS~20046$-$0623; IRAS~19252$-$7245:][]{ twmpifs,hos98},
but none are seen near the 400 km~s\( ^{-1} \) value as expected from a variety
of physical predictions.

One possibility deserves mention, which is that during the late-merger state,
eccentric orbits may conspire with the delay in coordinating the accumulation
of fuel for a nuclear starburst in such a way as to select ULIRGs at the apocenters
of their orbits, when velocities are low. Referring to Figure~\ref{fig:mihos},
during the final merger process (well after the initial encounter), the nuclear
orbits are very eccentric, owing to the removal of angular momentum, largely
by tidal debris. The very high velocity peaks near 500 km~s\( ^{-1} \) are
associated with the close passages. If the ultraluminous activity is episodic
in nature, with relatively short duration bursts occurring in reaction to transitory
tidal disruptions, then the ULIRGs we see may represent stages shortly after
these close encounters. A finite time is required to organize gas motions in
response to the close encounter, with the ultraluminous activity turning on
only after a significant gas concentration has been established. It is not difficult
to imagine in this scenario that the gas---moving more slowly than the galaxy,
but with less distance to travel---will reach the nucleus at approximately the
same time that the galaxies reach apocenter. Keeping in mind that we are discussing
the events occurring after the second close encounter, the orbital timescale
is \( \sim 10^{8} \) yr, such that apocenter is reached a few\( \times 10^{7} \)
yr after close approach. This timescale is in rough agreement with expectations
of the dynamical timescale within a few kpc of the nucleus---the region from
which the fuel for the nuclear event is gathered. During the final merger process,
it is possible that several ultraluminous bursts occur following close interactions
of the nuclei. The idea of repetitive starbursts in the final stages of merging
was first explored by \citet{noguchi}, though in the context of cloud-cloud
collisions in this case. The scheme proposed here of delayed---and potentially
episodic---bursting would bias late-merger ULIRGs to larger separations and
therefore lower velocities, quite possibly consistent with the observed distribution.
The problem with first encounter ULIRGs remains, since the average first encounter
velocity is around 475 km~s\( ^{-1} \). Unfortunately the current sample is
not large enough to probe possible differences in the velocity distributions
of early and late ULIRG encounters.

\section{Summary}

Near-infrared spectra in the 2 \( \mu  \)m window have been obtained for a
nearly complete, volume-limited sample of 33 ULIRGs. Lines of atomic hydrogen
and helium recombination, molecular hydrogen vibration-rotation, {[}\ion{Fe}{2}{]},
and {[}\ion{Si}{6}{]} are seen in the spectra. The Pa\( \alpha  \) is predominantly
seen as a nuclear event, with the majority of the emission arising in the central
kpc of most ULIRGs. Initiated as a search for optically hidden active galactic
nuclei, this survey reveals only two such galaxies, both of which are characterized
as Seyfert galaxies based on visible-light spectra. It is found via comparison
of the Pa\( \alpha  \) and Br\( \gamma  \) lines that significant extinction
of several magnitudes persists into the near-infrared bands, such that buried
active nuclei can not be ruled out in many objects. It is estimated that with
the sensitivity of the current survey, at least 25 mag of visual extinction
would be needed to hide a QSO with half the bolometric luminosity of the ULIRG.

Strong molecular hydrogen emission is seen in all of the primary nuclear spectra.
The molecular hydrogen spectrum indicates that most H\( _{2} \) emission in
ULIRGs stems from thermal excitation rather than fluorescent processes. Some
ULIRGs appear to have a small fraction of their molecular hydrogen emission
owing to fluorescence, though none approach pure fluorescence.

The relative radial velocities found in this sample of ULIRGs does not exceed
\( \sim 120 \) km~s\( ^{-1} \), with an inferred maximum deprojected velocity
differential of around 200 km~s\( ^{-1} \). This relatively low velocity differential
is difficult to produce from interactions between galaxies with masses comparable
to that of the Milky Way using simple physical assumptions. It is suggested
that ULIRGs in the late stages of merging may be selectively found near the
time of apocenter, leading to a lower average relative velocity between nuclei.

\acknowledgements{}

We thank Michael Strauss for his role in defining the sample of ULIRGs, and
for participating in the early stages of the Caltech effort in studying the
2 Jy ULIRGs. We also thank Chris Mihos, Gerry Neugebauer, Nick Scoville, and
Sterl Phinney for helpful discussions. James Graham provided information on
{[}\ion{Fe}{2}{]} transition strengths and wavelengths. Many persons accompanied
us on the observing runs to Palomar, most notably Rob Knop and James Larkin,
both of whom provided expert tutelage on the use of the spectrograph, and on
methods of data reduction. We thank the night assistants at Palomar, Juan Carasco,
Rick Burruss, Skip Staples, and Karl Dunscombe for their assistance in the observations.
This research has made use of the NASA/IPAC Extragalactic Database (NED), which
is operated by the Jet Propulsion Laboratory, Caltech under contract with NASA.
T.W.M. is supported by the NASA Graduate Student Researchers Program, and the
Lewis Kingsley Foundation. This research is supported by a grant from the National
Science Foundation.

\appendix{}

\section{Procedures for Measuring Line Properties\label{procedures}}

The line properties reported in Tables~\ref{tab:mofoprim} and \ref{tab:mofosec}
represent measured values from the spectra, rather than Gaussian fit parameters.
Especially for the weaker lines, the quality of the continuum subtraction is
vital to proper line characterization. This is done with a simultaneous multiple
Gaussian line fit and quadratic continuum fit. In all but a few cases, this
results in a very close fit to the continuum. For those few cases where the
continuum fit is questionable, some of the weaker lines are excluded from Tables~\ref{tab:mofoprim}
and \ref{tab:mofosec}. All of the fitting procedures make use of the noise estimate
computed on a pixel-by-pixel basis, as described earlier. In the end, the noise
is scaled slightly (typically by factors of less than 30\%) in order to make
the final reduced chi-squared value equal to one. The scaled noise is used in
estimating errors on the values in Tables~\ref{tab:mofoprim} and \ref{tab:mofosec}.
Each of the line parameters is computed based on the following line definition
procedure. Each potential line is searched for by summing two resolution elements
(7 pixels) of information around the line center, and determining if there is
emission above the continuum exceeding 2.5\( \sigma  \). For lines satisfying
this condition, cutoffs are established on the red and blue sides of the line
center by the following rules:

\begin{enumerate}
\item If the spectrum extends below the continuum level, establish a cutoff, including
only positive valued points.
\item If there exists a local minimum, followed by at least three higher points, assume
that the following points belong to another line and exclude them, keeping only
the minimum point.
\item If the next two pixels are between 0--1\( \sigma  \) above the continuum, include
them and quit.
\end{enumerate}
These procedures obviously can not practice the same discrimination that a diligent
scientist might, but we find that a reasonable job is performed---especially
on lines whose boundaries are obvious to the eye. Nonetheless, lines with a
final computed equivalent width measuring less than 2.5\( \sigma  \) are typically
excluded, though those that appear to be real in the one-dimensional spectra
are preserved. Conversely, a few lines with reported detections greater than
2.5\( \sigma  \) are excluded on the basis of either poor local continuum fit
or simply less-than-believable results.

Because the weaker lines do not have a well defined full-width at half-maximum,
the FWHM values in Tables~\ref{tab:mofoprim} and \ref{tab:mofosec} are based
on a combination of line flux and peak amplitude. If a Gaussian shape is assumed,
the FWHM of the line would be equal to the line flux divided by the peak amplitude,
times the proportionality constant of \( \sqrt{\ln (256)/2\pi }=0.94 \). Performed
in this manner, it is relatively simple to compute the error of such an estimate.
A strict measure of the FWHM is provided in Tables~\ref{tab:mofoprim} and \ref{tab:mofosec}
for the Pa\( \alpha  \) line in each galaxy for comparison to the value estimated
by this technique. All of the reported FWHM values have been reduced to intrinsic
line widths by subtracting the instrumental resolution in quadrature. Each spectrum
uses the measured line widths in the wavelength calibration spectrum as a basis
for the decomposition. Values in Tables~\ref{tab:mofoprim} and \ref{tab:mofosec}
for which the measured line width is exceeded by the instrumental width are
listed as unresolved lines of zero width. Almost all of the lines in this survey
are resolved, typically around 200--300 km~s\( ^{-1} \) in width.

The velocity measurements in Tables~\ref{tab:mofoprim} and \ref{tab:mofosec}
are indicated relative to Pa\( \alpha  \), though the Pa\( \alpha  \) velocity
itself is expressed as a direct quantity in km~s\( ^{-1} \). These values
are simply computed as first-moment measures in the spectral lines, with error
estimates computed from the noise estimates in a straightforward way. Added
to this error estimate is a wavelength calibration error, assumed to be 10--15
km~s\( ^{-1} \), and a slit placement error. The latter uncertainty attempts
to estimate the wavelength offset induced by a non-uniform slit illumination.
When the line source is spatially compact enough to produce a discernible peak
somewhere in the slit, the location of this peak within the slit influences
the ultimate wavelength associated with this emission. For these purposes, it
is assumed that the line source has a real physical size roughly half that of
the continuum, which is almost always spatially resolved. Further, it is assumed
that the source is centered in the slit with a Gaussian 1\( \sigma  \) placement
error of one-quarter of a slit width. Only cases of exceptional seeing result
in an appreciable error from this source, with a median error estimate of 15
km~s\( ^{-1} \) and a maximum of 38 km~s\( ^{-1} \). Despite the accounting
of uncertainties, the velocities indicate a dispersion outside of the error
budgets that is not correlated with line species or grating setting. It may
well be that the variable spectral range over which the line is summed is largely
responsible for determining the offset, as the range is biased to contain only
positive points, and will extend as far as possible from the line center to
include such points. With the large moments generated from distant points, the
estimated velocity may wander far from its true center. Thus the values of velocity
in Tables~\ref{tab:mofoprim} and \ref{tab:mofosec}, especially for weaker lines,
are perhaps of little use. Gaussian fit centers may do a better job of appropriately
representing the line velocities.

\section{Estimating Extinction to Hidden AGN\label{qsoextinct}}

The median QSO spectrum is constructed from nine optically identified QSOs in
the redshift range \( 0.089<z<0.182 \), seven of which have available \( K \)
band photometry. Comparing the absolute \( K \) band magnitudes to bolometric
luminosities of the quasars, one finds the approximate relation \( M_{K}=4.0\pm 0.5-2.5\log \left( L_{bol}/L_{\odot }\right)  \).
The median infrared luminosity of the 31 non-AGN ULIRGs in the present sample
is \( 1.1\times 10^{12}\, L_{\odot } \). A QSO within a ULIRG that contributes
half of the bolometric luminosity---most of which is reprocessed into the infrared---would
then have \( L_{bol}=5.5\times 10^{11}\, L_{\odot } \), or \( M_{K}=-25.4\pm 0.5 \)
mag. This establishes the \( K \) band flux density of the QSO continuum. At
a rest wavelength of 1.881 \( \mu  \)m, corresponding to \( +1000 \) km~s\( ^{-1} \)
in velocity, it is found that the median QSO has an excess flux density due
to Pa\( \alpha  \) measuring 22\% of the baseline continuum level \citep{twm99}.
Thus the absolute magnitude of the line emission at this velocity is \( M_{K}(Pa\alpha _{1000})=-23.74\pm 0.5 \)
mag.

Typical ULIRG continuum levels can be approximated in a similar way as that
above, using 5\( '' \) aperture photometry of the sample galaxies where available
\citep{kimthesis,twmpifs}. Eight of the 31 non-AGN ULIRGs have \( K_{s} \)
measurements, and comparing to the infrared luminosity (8--1000 \( \mu  \)m)
leads to the relation \( M_{K_{s}}=6.3\pm 0.6-2.5\log (L_{IR}/L_{\odot }) \).
Therefore the typical non-AGN ULIRG in this sample, with a median luminosity
\( \log (L_{IR}/L_{\odot })=12.05 \), has \( M_{K_{s}}=-23.83\pm 0.6 \) mag.

The expected ratio of broad Pa\( \alpha  \) to the ULIRG continuum at 1.881
\( \mu  \)m is then 0.92, with the 1\( \sigma  \) range extending from 0.44
to 1.91. If a given ULIRG has a spectrum with a signal-to-noise ratio \( G \),
then diminishing the QSO Pa\( \alpha  \) to the 1\( \sigma  \) level requires\[
\frac{Q_{Pa\alpha }e^{-\tau }}{Q_{cont}e^{-\tau }+U_{cont}}=\frac{1}{G},\]
 where \( Q \) represents QSO contributions to continuum and line flux, \( U \)
is the ULIRG continuum contribution, and \( e^{-\tau } \) is the attenuating
factor. With the additional relations that \( Q_{Pa\alpha }/Q_{cont}=0.22 \)
and \( Q_{Pa\alpha }/U_{cont}=0.92 \), we find that the required attenuation
is \( e^{\tau }=0.92(G-1/0.22) \), or \( 2.5\log (0.92G-4.2) \) magnitudes
at Pa\( \alpha  \). The visual extinction is roughly 7 times this value.

\section{Molecular Hydrogen Diagnostics\label{append}}

\subsection{Population Diagrams}

The specific mechanism responsible for the H\( _{2} \) emission can be investigated
via population diagrams. By accounting for various degeneracies, transition
probabilities, and photon energy, one may compute the relative populations of
the upper states of each transition based on the observed line flux. In pure
thermal equilibrium, the population levels will follow an exponential profile
with respect to upper state energy. More formally, the intensity of a given
transition is described by\[
I(v_{1},v_{2},J_{1},J_{2})\propto g_{s}(J_{1})A(v_{1},v_{2},J_{1},J_{2})(2J_{1}+1)h\nu _{12}e^{-\frac{E_{1}}{kT}},\]
where the upper and lower states are denoted by the 1 and 2 subscripts, \( v \)
represents vibration state, \( J \) the rotation state, \( g_{s}(J_{1}) \)
the ortho/para degeneracy, \( A(v_{1},v_{2},J_{1},J_{2}) \) the transition
probability, \( \nu _{12} \) the emitted photon frequency, \( E_{1} \) the
upper state energy, and \( T \) the equilibrium temperature. Therefore a logarithmic
plot of the line intensities divided by the terms multiplying the exponential
above produces a linear relation against the upper energy scale with a slope
proportional to the inverse of the temperature. The physical parameters for
selected \( K \) band transitions are given in Table~\ref{tab:h2phys}.

\begin{deluxetable}{lcccc} 
\tabletypesize{\footnotesize}
\tablewidth{0pt}
\tablecaption{Physical Parameters of H$_2$ Transitions\label{tab:h2phys}}
\tablehead{\colhead{Transition} & \colhead{$\lambda_{air}$ } &\colhead{$J_1$} & \colhead{$A(v_1,v_2,J_1,J_2)$\tablenotemark{a}} & \colhead{$E_1/k$\tablenotemark{b}}\\ 
& \colhead{($\mu$m)} & & \colhead{($10^{-7}$ s$^{-1}$)} & \colhead{(K)}}
\startdata 

1--0 S(0) & 2.22268 & 2 & 2.53 & \phn 6471 \\

1--0 S(1) & 2.12125 & 3 & 3.47 & \phn 6951 \\

1--0 S(2) & 2.03320 & 4 & 3.98 & \phn 7584 \\

1--0 S(3) & 1.95702 & 5 & 4.21 & \phn 8365 \\

1--0 S(4) & 1.8914\phn & 6 & 4.19 & \phn 9286 \\

1--0 S(5) & 1.8353\phn & 7 & 3.96 & 10341 \\[5pt]

2--1 S(0) & 2.3550\phn & 2 & 3.68 & 12095 \\

2--1 S(1) & 2.2471\phn & 3 & 4.98 & 12550 \\

2--1 S(2) & 2.1536\phn & 4 & 5.60 & 13150 \\

2--1 S(3) & 2.0729\phn & 5 & 5.77 & 13890 \\

2--1 S(4) & 2.0035\phn & 6 & 5.57 & 14764 \\

2--1 S(5) & 1.9443\phn & 7 & 5.05 & 15763 \\

2--1 S(6) & 1.8942\phn & 8 & 4.30 & 16880 \\

2--1 S(7) & 1.8523\phn & 9 & 3.38 & 18107 \\[5pt]

3--2 S(3) & 2.2008\phn & 5 & 5.63 & 19085 \\

\enddata
\tablenotetext{a}{Taken from \citet{turner}.}
\tablenotetext{b}{Taken from \citet{dabrowski}.}
\end{deluxetable}

When the molecular hydrogen is not in thermal equilibrium---such as may be the
case in fluorescent excitation by ultraviolet photons---a diagram such as that
described above will not show all points lying along a common line, but rather
disconnected linear segments, with common vibration states each defining a line
connecting the different rotation states. The characteristic temperature of
this line is referred to as the rotation temperature, \( T_{rot} \). Ideally,
each vibration state has the same characteristic rotation temperature, and these
segments can be connected with another line characterized by a vibration temperature,
\( T_{vib} \). The thermal case simply has \( T_{rot}=T_{vib}=T \). Typically,
pure fluorescent conditions are characterized by \( T_{rot}=800 \)---1400 K,
and \( T_{vib}=5000 \)--6000 K \citep{bvd,tso}. Excitation conditions associated
with thermal processes are typically found to have \( T_{rot}=1600 \)--2400
K, with a slight difference between rotation and vibration temperatures characterized
by \( T_{rot}\approx 0.8T_{vib} \) \citep{tanaka,hasegawa,beckwith}.

\subsection{Typical \protect\( K\protect \) Band H\protect\( _{2}\protect \) Line Ratios}

Because ULIRGs tend to show relatively uniform H\( _{2} \) emission properties---namely
those corresponding to thermal emission at 1500--2500 K---it may be useful for
other researchers to have a table of expected H\( _{2} \) line strengths in
the \( K \) band for a range of typical temperatures. Table~\ref{tab:h2} provides
such a reference, with line intensities normalized to the H\( _{2} \)~1--0
S(1) line. These estimates assume an ortho-to-para ratio of 3:1, and \( T_{rot}=0.8T_{vib} \),
as is found empirically \citep{tanaka,hasegawa}. In order to regain the pure
thermal regime characterized by a single temperature, the 2--1 entries in Table~\ref{tab:h2}
should be multiplied by \( \exp (-1107/T) \), and the 3--1 transition multiplied
by \( \exp (-2144/T) \), lowering these values considerably.

\begin{deluxetable}{lccccc} 
\tabletypesize{\footnotesize}
\tablewidth{0pt}
\tablecaption{Expected H$_2$ Line Strengths Relative to H$_2$ 1--0 S(1)\label{tab:h2}}
\tablehead{\colhead{Transition} & \colhead{$\lambda_{air}$ ($\mu$m)} &\colhead{$T_{rot}=1500$ K} & \colhead{$T_{rot}=2000$ K} & \colhead{$T_{rot}=2500$ K} & \colhead{$T_{rot}=4000$ K} \\ } 
\startdata 

1--0 S(0) & 2.22268 & 0.228 & 0.211 & 0.201 & 0.187 \\

1--0 S(1) & 2.12125 & 1.000 & 1.000 & 1.000 & 1.000 \\

1--0 S(2) & 2.03320 & 0.336 & 0.374 & 0.398 & 0.438 \\

1--0 S(3) & 1.95702 & 0.805 & 1.019 & 1.174 & 1.450 \\

1--0 S(4) & 1.8914\phn & 0.177 & 0.261 & 0.329 & 0.468 \\

1--0 S(5) & 1.8353\phn & 0.295 & 0.519 & 0.728 & 1.211 \\[5pt]

2--1 S(0) & 2.3550\phn & 0.015 & 0.030 & 0.045 & 0.083 \\

2--1 S(1) & 2.2471\phn & 0.068 & 0.143 & 0.225 & 0.441 \\

2--1 S(2) & 2.1536\phn & 0.023 & 0.053 & 0.089 & 0.191 \\

2--1 S(3) & 2.0729\phn & 0.055 & 0.145 & 0.259 & 0.622 \\

2--1 S(4) & 2.0035\phn & 0.012 & 0.037 & 0.072 & 0.197 \\

2--1 S(5) & 1.9443\phn & 0.020 & 0.072 & 0.156 & 0.496 \\

2--1 S(6) & 1.8942\phn & 0.003 & 0.014 & 0.033 & 0.124 \\

2--1 S(7) & 1.8523\phn & 0.004 & 0.020 & 0.054 & 0.246 \\[5pt]

3--2 S(3) & 2.2008\phn & 0.003 & 0.016 & 0.045 & 0.202 \\

\enddata
\tablecomments{This table assumes that $T_{rot}=0.8T_{vib}$. To recover a strictly thermal regime ($T_{rot}=T_{vib}=T$), multiply the 2--1 transistions by $\exp (-1107/T)$ and the 3--2 transition by $\exp (-2144/T)$.}
\end{deluxetable}

\end{document}